\documentclass{aa}

\usepackage{graphicx}
\usepackage{subcaption}
\usepackage{lscape}
\usepackage{placeins}
\RequirePackage{fix-cm}
\usepackage{amsmath}
\usepackage{bm}
\usepackage{newtxtext,newtxmath}
\usepackage[T1]{fontenc}
\usepackage{xspace}
\usepackage[hidelinks]{hyperref}
\usepackage{cleveref}
\usepackage{chngcntr}
\usepackage{multirow}
\usepackage{tabularx}
\usepackage{natbib}
\bibpunct{(}{)}{;}{a}{}{,}

\defcitealias{pavan_jet-environment_2023}{P23}
\defcitealias{pavan_role_2025}{P25}

\newcommand\s{\mathrm{s}}

\newcommand\m{\mathrm{m}}
\newcommand\cm{\mathrm{cm}}
\newcommand\km{\mathrm{km}}

\newcommand\g{\mathrm{g}}
\newcommand\msol{\mathrm{M}_\odot}
\newcommand{\msun}{\mathrm{M}_\odot}

\newcommand\erg{\mathrm{erg}}

\newcommand\rad{\mathrm{rad}}

\newcommand\G{{\rm G}}

\newcommand\obs{\mathrm{obs}}
\newcommand\tot{\mathrm{tot}}
\newcommand\iso{\mathrm{iso}}
\newcommand\sh{\mathrm{sh}}
\newcommand\mprot{m_\mathrm{p}}
\newcommand\kboltz{k_\mathrm{B}}
\newcommand\mw{\mu_\mathrm{mw}}
\newcommand{\bo}{\mathrm{bo}}

\newcommand{\jet}{\mathrm{j}}

\newcommand\ej{\mathrm{ej}}
\newcommand{\beam}{\mathrm{beam}} 
\newcommand{\floor}{\mathrm{floor}}
\newcommand{\atmo}{\mathrm{atmo}}

\newcommand{\delay}{\mathrm{GW-EM}}

\newcommand{\thresh}{\mathrm{thresh}}

\newcommand{\de}{\mathrm{d}}
\newcommand{\di}{\partial}
\renewcommand{\vec}[1]{\boldsymbol{\mathbf{#1}}}
\newcommand{\vers}[1]{\hat{\vec{#1}}}
\renewcommand{\epsilon}{\varepsilon}

\begin{document}
   \title{Modelling the delayed shock-breakout emission \\following jet-launching binary neutron star mergers \\via relativistic magnetohydrodynamic simulations}

\titlerunning{Modelling the delayed shock-breakout emission of jets in BNS via relativistic MHD simulations}

\author{Matteo Pais\inst{1,2}\corrauth{matteo.pais@inaf.it}\and
Riccardo Ciolfi\inst{1,2}\email{riccardo.ciolfi@inaf.it}\and
Andrea Pavan\inst{1,2}\email{andrea.pavan@inaf.it}
}

\institute{INAF, Osservatorio Astronomico di Padova, Vicolo dell'Osservatorio 5, I-35122, Padova, Italy \and INFN, Sezione di Padova, Via Francesco Marzolo 8, I-35131 Padova, Italy}
\date{Received DDMM, 2026}

\abstract
{In binary neutron star (BNS) mergers launching a relativistic jet, an electromagnetic (EM) signal is produced when the jet-driven shock breaks out of the merger ejecta. The observed time delay of this shock-breakout (SBO) emission with respect to the gravitational-wave (GW) signal from the merger provides a powerful probe of the physical conditions governing jet launching and early-time jet propagation. }
{Considering different models of jet propagation in realistic post-merger environments, we investigate the SBO emission and corresponding GW-EM delay that would be observed depending on the viewing angle and the assumed ejecta opacity. } 
{We performed relativistic magneto-hydrodynamic simulations of jets propagating through a post-merger environment directly imported from the outcome of a previous BNS merger simulation. We also introduced a specific procedure to faithfully reconstruct the early dynamical ejecta up to their natural front. The evolution was followed in 3D up to 0.6\,s and then we continued imposing axisymmetry and an eight times higher resolution. Varying jet launching time and luminosity, we identified three representative models spanning regimes from early breakout to extended jet choking. For each case, we tracked the jet-driven forward shock up to the photosphere and computed the angle-dependent bolometric SBO luminosity, assuming full conversion of the thermal energy within the shocked material into radiation, and taking into account non-radial photon propagation, relativistic Doppler shifts, and light-travel-time effects. We considered two opacity values spanning a factor of ten. }
{We find that the GW-EM delay depends only weakly on both the viewing angle and the ejecta opacity, making it, at least for the limited set of configurations considered here, a robust diagnostic for constraining models. Comparing our three cases with GRB 170817A, we find smaller GW-EM delays. Reproducing the same delay will require broader exploration of the parameter space. }
{}


\maketitle
\nolinenumbers
\section{Introduction}
\label{sec: intro}

The detection of gravitational waves (GWs) and electromagnetic (EM) emission from the binary neutron star (BNS) merger GW170817 marked the beginning of multi-messenger astrophysics with GW sources \citep{LVC-GRB, LVC-BNS,LVC-MMA}.
The EM counterparts of the event included a short gamma-ray burst (GRB) and a kilonova, the latter being powered by the radioactive decay of heavy elements synthesized in the merger ejecta (\citealt{Metzger2020}). 
Together, these observations offer a unique opportunity to probe the physics of compact binary mergers, relativistic outflows, and $r$-process nucleosynthesis.

The prompt gamma-ray signal accompanying the event, named GRB\,170817A, was orders of magnitude less luminous than typical short GRBs, leaving initial doubts on its true origin \citep{LVC-GRB}. 
Later on, long-term observations of the afterglow signal confirmed the presence of a highly collimated, relativistic jet compatible with a short GRB, but pointing $\approx\!15-30$ degrees away from the line of sight \citep{mooley_superluminal_2018, Ghirlanda2019}. 
This consolidated the idea that GRB\,170817A was actually produced by a lateral portion of the outflow with a limited Lorentz factor ($<10$). 
A promising mechanism invoked to explain this high-energy emission, which we adopt in the present work, is the shock breakout associated with the expanding jet-cocoon system (\citealt{kasliwal_illuminating_2017, gottlieb_cocoon_2018}; see \citealt{nakar_early_2010, nakar_relativistic_2012} for the theory of (relativistic) shock breakout, \citealt{bromberg_-rays_2018} for relativistic magnetohydrodynamic shock-breakout or cocoon emission in the context of GW170817, and \citealt{irwin_propagation_2019} for the shape and propagation of cocoons in an ambient medium).

In this context, a key observable to understand the post-merger evolution is the measured $\approx\!1.74$\,s delay between the GW signal and the onset of the gamma-ray emission (\citealt{LVC-GRB}; see \citealt{gill_when_2019} and \citealt{zhang_delay_2019} for discussions on the physical origin of this GW-EM delay).
While a first obvious contribution to this delay is the time that elapsed between the merger and the launch of the relativistic jet, additional delay accumulates as the incipient jet gradually accelerates towards ultra-relativistic velocities while piercing through the surrounding medium. 
In the process, a jet-cocoon system is formed and its front shock propagates through the environment, until this environment becomes too diluted to keep photons trapped within the shock itself \citep{nakar_electromagnetic_2020}. 
The time and radial distance characterizing the shock breakout depend on the angular distance from the jet injection axis.  
Moreover, the opacity of the ambient material can play an important role, as it defines the breakout condition. 

When the jet is launched by the central engine, the first obstacle is represented by the dense material expelled in the post-merger phase up to the jet launch time. 
The mass outflow in this phase is dictated by the high temperatures reached during the merger, which is enhanced by the growing magnetic fields with the additional contribution of neutrino heating \cite[and refs. therein]{foucart_neutrino_2023}. 
The time it takes for the jet to be launched is a critical parameter of the problem, regulating how massive and extended this post-merger outflow component can be (e.g. \citealt{pavan_role_2025}). 
Beyond this component, a second obstacle is represented by the material that is ejected dynamically during the merging process, characterized by a radially decreasing density profile and increasing velocity (e.g. \citealt{bauswein_systematics_2013, hotokezaka_mass_2013}). 
The bulk of the dynamical ejecta can reach velocities as high as $\approx\!0.8\,c$ (e.g. \citealt{radice_viscous-dynamical_2018}), surrounded by an even faster, extremely low density layer with a dynamically irrelevant mass and energy.
While the final properties of the escaping jet are mostly determined by the interaction with the denser post-merger outflows, a proper description of the bulk of the dynamical ejecta up to the fast outer tail becomes crucial in the investigation of the angle-dependent GW-EM delay, since the corresponding optical depth can strongly affect the exact time and radius of the shock breakout (e.g. \citealt{Gutierrez2025}).\footnote{Capturing the evolution of dynamical ejecta up to their outer front is a challenge in BNS merger simulations, due to the relatively high numerical density floor employed.
Only recently, the use of a much lower density floor with rapidly decreasing radial profile has proven effective in overcoming the problem (\citealt{kalinani_jet-environment_2025}, and refs.~therein).}

In the present work, we investigate the origin of the GW-EM delay in GW170817-like BNS mergers, by means of special relativistic magnetohydrodynamic (RMHD) simulations of incipient jets that were launched at a given time after merger and are propagating through realistic post-merger environments.
These environments are based on the results of general relativistic magnetohydrodynamic (GRMHD) simulations of BNS mergers by \citet{ciolfi_collimated_2020} and combine the directly imported post-merger outflow component with dynamical ejecta faithfully reconstructed up to the fast outer tail from the original mass outflows at a radial distance of 300\,km.\footnote{Here by `realistic' environments we refer to the fact that they are directly imported from BNS merger simulations. On the other hand, the jet-launching prescription is imposed manually. This choice does not allow for a direct physical correlation between the jet-launching time, jet power, and surrounding ejecta mass expected in central-engine models \citep{Gottlieb_unified_2025}; a self-consistent coupling of the jet and environment is left to future work.}
The evolution of the jet-environment system and the propagation of the jet-driven shock was initially followed in 3D, and then continued in 2D by imposing axisymmetry. 
By monitoring the optical depth ahead of the shock front, we established when the breakout condition was satisfied and thus determined the resulting GW-EM delay depending on the angular distance from the jet injection axis and the opacity of the ambient medium. 
Furthermore, we computed the isotropic-equivalent bolometric luminosity of the shock-breakout emission, assuming that the thermal energy within the shock has been fully converted into observable radiation. 
We explored the effects of varying the physical model in terms of the jet launch time and luminosity and compared our results with the properties of GRB\,170817A, highlighting the potential of the GW-EM delay as a diagnostic of jet propagation scenarios.

The paper is organized as follows. 
In Sect.~\ref{sec: methods}, we describe the numerical setup and methodology, including the construction of the initial conditions for the environment, the jet-launching prescription, and the extension of the evolution in 2D after the jet engine activity was switched off (see below). The same section also presents our different models and the prescription adopted to identify the shock front and the breakout condition.
Sect.~\ref{sec: evolution_results} describes the dynamical evolution of the system, while in Sect.~\ref{sec: gw_em_delay} we present in detail the results in terms of GW-EM delay and isotropic-equivalent bolometric luminosity, also comparing them with GW170817 and GRB\,170817A. 
Finally, Sect.~\ref{sec: conclusions} summarizes our main findings and elaborates on their implications for interpreting GW-EM delays in future multi-messenger observations.

\section{Simulation setup and methods}
\label{sec: methods}

\begin{figure*}
    \centering
    \includegraphics[width=0.98\linewidth]{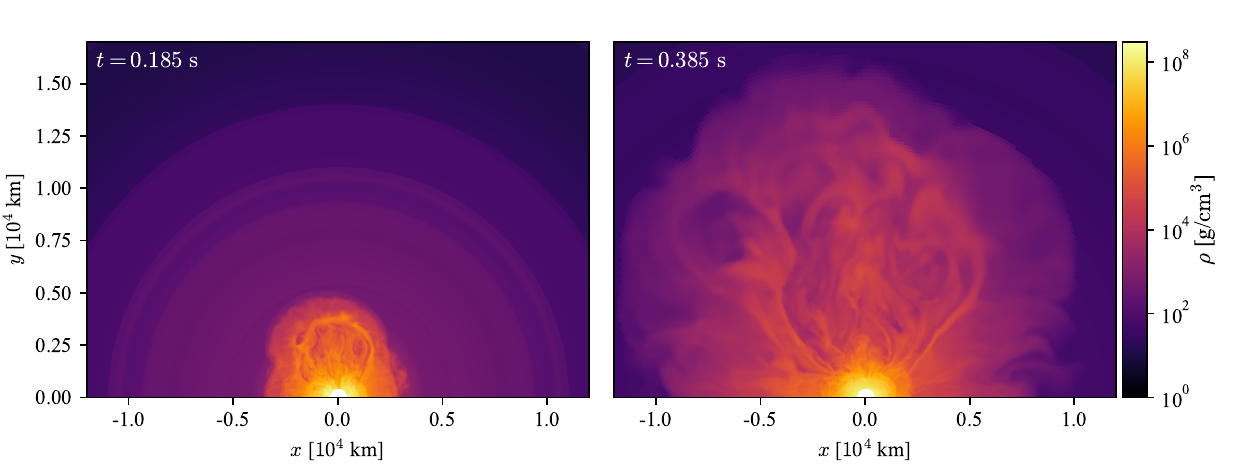}
    \vskip -0.3cm
    \caption{Meridional view of rest-mass density at the time of jet injection for model A (left) and for models B and C (right), namely at $185$\,ms and 385\,ms after merger, respectively.}
    \label{fig: rho_IC}
\end{figure*}

Our study is based on simulations performed with the publicly available RMHD module of the {\textsc{pluto}} code \citep[v4.4-patch3,][]{mignone_pluto_2007}, following two sequential steps. 
First, we injected a relativistic jet into a realistic post-merger environment, directly imported from the outcome of a GRMHD BNS merger simulation (see Sect.~\ref{sec: initialdata}), and evolved the system in full 3D.
During this initial phase, we followed the jet-environment interaction up to $t = 0.6$\,s after jet launch, corresponding to twice the characteristic decay timescale of the jet injection luminosity (Sect.~\ref{sec: jetprescription}). 
At this point, the jet has crossed the densest regions of the environment, consisting of the post-merger outflows driven by the massive neutron star (MNS) remnant at the centre. 
To track the subsequent evolution over an order of magnitude longer timescale, we constructed an azimuthal average of the system weighted by the energy density and continued evolving under the assumption of axial symmetry (Sect.~\ref{sec: axisym}).
The resulting 2D description allows us to dramatically increase the resolution while extending the radial domain up to the scales where the flow approaches homologous expansion ($t \gtrsim 10-20$\,s). 

In the following subsections, we describe in detail the initial data and 3D grid setup (Sect.~\ref{sec: initialdata}), the jet injection prescription (Sect.~\ref{sec: jetprescription}), our different models and the numerical methods employed for the evolution (Sect.~\ref{sec: models-methods}), and the procedure used to continue the evolution in 2D (Sect.~\ref{sec: axisym}).
Finally, we discuss the methods used to follow the propagation of the shock front and to establish the angle-dependent breakout condition (Sect.~\ref{sec: Shockfront_BO}).

\subsection{Initial data and grid setup}
\label{sec: initialdata}

Following \citet{pavan_short_2021, pavan_jet-environment_2023, pavan_role_2025}, the 3D simulations were carried out in spherical coordinates $(r,\theta,\phi)$ with an inner excised region of radius $r_0 = 380~\km$, outside of which general relativistic effects can be neglected, and with the original orbital axis of the BNS, corresponding to jet injection axis, oriented orthogonally to the coordinate polar axis.
The tilted coordinates $(r,\Theta,\Phi)$ for which the corresponding polar axis is instead aligned with the jet injection axis are linked to the spherical coordinates of the simulation ($r,\theta,\phi$) via the expressions reported in Appendix~\ref{appendix: tilted_reference_frame}.

The radial domain extended up to $r_{\max} = 2.5\times10^6~\km$, with logarithmic spacing in $r$ and uniform angular spacing in $\theta \in [0.1,\pi-0.1]~\rad$ and $\phi \in [0,\pi]$. Note that we evolved only half of the full 3D domain, corresponding to the northern side of the system, as in the reference BNS merger simulation.
The extension of the domain along $\theta$ was chosen to avoid the polar axis singularity. 
The resolution was set to $N_r \times N_\theta \times N_\phi = 768 \times 256 \times 256$, yielding $\Delta r\approx r\Delta\theta \approx r\Delta\phi \approx 4.4$\,km at the excision radius ($r = 380$\,km).

We considered two initial environments where the jet was injected, based on the same ones employed in \citet{pavan_role_2025} (hereafter P25), which are in turn built on a GRMHD BNS~merger simulation of \citet{ciolfi_collimated_2020}. The corresponding jet injection times are $t_\jet=185$\,ms and $385$\,ms after merger, respectively. 

The merging system of \citet{ciolfi_collimated_2020} has a chirp mass matching GW170817, with $1.44$ and $1.29~\msun$ masses (corresponding to a mass ratio $q\simeq0.9$), and is modelled adopting the APR4 EoS (as implemented in \citealt{Endrizzi2016}), without including neutrino transport. The simulation has a resolution with finest grid spacing of $250~\m$, and a computational domain extending up to $3400~\km$ along all axes (equatorial symmetry is imposed on the orbital plane).
The resulting magnetized post-merger baryon wind carries a mass of order $\sim\!0.1~\msun$ and expands with a maximum velocity of $\approx\!0.2\,c$ (see \citealt{CiolfiKalinani2020}).

The outer environment, outside the front of the MNS-driven outflows, consists of material dynamically expelled during the merger process, i.e.~the dynamical ejecta.
Since the original BNS merger simulation employed a uniform and rather high density floor ($6.3\times10^4\,\mathrm{g\,cm^{-3}}$), such ejecta were suppressed above distances of $\sim\!1000$\,km. 
In order to include a faithful description of dynamical ejecta up to the fast low-density tail at their front, which is a key ingredient in the present investigation, we reconstructed this component consistently with the merger simulation, employing the original matter outflows up to a reliable distance of 300\,km. 
This reconstruction, detailed in Appendix~\ref{appendix: dynamical ejecta}, took input from the matter profiles along the orbital axis (positive $z$-axis of the merger simulation) and then assumed spherical symmetry of the dynamical ejecta. For the late-time evolution of the jet, we are indeed mostly interested in reproducing well the ejecta distribution around the orbital axis. 
For reconstructing the dynamical ejecta above 300\,km distance, we simulated their expansion substituting the original high-density floor with a much lower one, having radial profile decreasing as $r^{-6.5}$. This choice has proven effective in removing floor effects on the jet evolution up to very large scales \citep{pavan_jet-environment_2023}.

The initial environments adopted in our 3D simulations were thus the same employed in P25 up to the maximum extension of the MNS-driven outflows (along each individual angular direction), while in the outer part we substituted the original data with the reconstructed dynamical ejecta surrounded by the very tenuous material corresponding to the new density floor decreasing as $r^{-6.5}$.
The transition, occurring along each direction at the radius where the original density becomes smaller than the one of the reconstructed dynamical ejecta, is smooth everywhere and ensures continuity in density, pressure, and velocity.
The resulting density distributions of the two initial environments in the meridional plane are shown in Figure~\ref{fig: rho_IC}. We refer to Appendix~\ref{appendix: dynamical ejecta} and Figure~\ref{fig: ejecta 185ms and 385ms} for 1D radial profiles of density, velocity, and pressure of the reconstructed dynamical ejecta.

We summarize the mass and energy budget of the adopted environment (northern hemisphere only), separating the dense post-merger outflow (`inner'), the reconstructed dynamical ejecta, and the residual floor-level material (`numerical tail'). 
At 385 ms (185 ms), the mass and energy are dominated by the inner outflow with $M\simeq8.75~(7.8)\times10^{-2}~\msun$ and $E\simeq2.7~(2.54)\times10^{50}$\,erg, while the dynamical ejecta contribute with $M_{\rm dyn}\simeq5~(5.9)\times10^{-4}~\msun$ and $E_{\rm dyn}\simeq3~(3.2)\times10^{49}$\,erg. The numerical tail is dynamically and energetically negligible, with a mass of $10^{-9}~\msun$. 
The reconstructed dynamical ejecta account for the material beyond the inner outflow front, at radii above $\simeq\!1.25\times10^4$\,km ($\simeq\!4000~$\,km) at 385~ms (185~ms) after the merger (see Appendix~\ref{appendix: dynamical ejecta} for the profiles).

Although broadly consistent with the range of ejecta properties inferred for GW170817, the adopted environments are not fine-tuned to reproduce that event. The total unbound mass of several times $10^{-2}~\msol$, dominated by the post-merger outflow, lies at the upper end of the $\sim (2-5)\times10^{-2}~\msol$ inferred for GW170817 (e.g.~\citealt{Metzger2020}).

The environment is aspherical. The mass per unit solid angle increases from $\sim1.24~(0.6)$ to $\sim1.69 ~(1.7)\times10^{-2}~\msol$~\,sr$^{-1}$ between the pole and the equator, whereas the energy per unit solid angle decreases from $\sim 8.2 ~(9.9) \times~10^{49}$ to $\sim4.7~(4.5)\times10^{49}$\,erg\,sr$^{-1}$. 
The radial velocity increases outwards, reaching $\approx0.15$--$0.2\,c$ within the inner outflow and increasing up to $v_{\rm ej}\simeq0.8\,c$ at the outer edge of the bulk of the dynamical ejecta (cf.~Eq.~\ref{eq: r_de_last}), which sets the shock-breakout condition.

\subsection{Jet injection prescription}
\label{sec: jetprescription}

The jet was injected at the excision radius $r_0$ within a fixed half-opening angle of $\theta_\jet\!=\!10^\circ$, following the prescription described in Appendix~\ref{appendix: jet}.
The one-sided jet luminosity, defined as the integral of the radial energy flux (matter enthalpy plus Poynting flux) over the injection cap \citep[as in][]{pavan_role_2025}, is 
\begin{equation}
\label{eq: luminosity}
    \begin{aligned}
    L_\jet = &2\pi r_0^2 \int_0^{\theta_\jet} \left[ \rho_\jet h^*_\jet  c^2 \Gamma^2_{\jet}  \beta^r_{~\jet} \right. \\ & \left.- B^r_{~\jet} (\vec{\beta}_{\jet}\cdot \vec{B}_\jet) - \Gamma^2_{\jet} (\vec{\beta}_{\jet}\cdot \vec{B}_\jet)^2 \beta^r_{~\jet} \right] c \sin\Theta\,\de\Theta \, ,
    \end{aligned}
\end{equation}
where $\vec{\beta}$, $\vec{B}$, and $\rho$ are velocity in units of $c$, magnetic field, and rest-mass density, respectively. 
The subscript `\,$\jet$\,' denotes quantities associated with the jet injection.
In the above expression, we imposed a constant and $\Theta$-independent Lorentz factor $\Gamma_\jet=3$ and $h^*_{\mathrm{j}} = (1+\sigma_\jet)h_\jet = 100$, where $h_\jet$ is the specific (dimensionless) enthalpy and $\sigma_\jet=b^2/(\rho h_\jet c^2)$ the local (comoving) magnetization (see Appendix~\ref{appendix: jet} for the exact definitions and profiles).
The choice $h^*_\jet=100$ fixes the total (thermal plus magnetic) energy per unit rest mass injected in the jet, ensuring a GRB-compatible terminal Lorentz factor of 300 at injection; with the field normalization adopted here (see below) the jet is weakly magnetized ($\sigma_\jet\!\ll\!1$), so $h^*_\jet\simeq h_\jet$. We note that this reflects the injection properties at 380\,km radius, which can still be consistent with a jet that has much larger magnetization at base (i.e.~close to the central engine).
The corresponding transverse profiles of $\sigma_\jet$ and $h_\jet$ are nearly top-hat (see Fig.~\ref{fig: transverse_equilibrium}).
In the tilted reference frame, the jet has a radial and azimuthal magnetic field, with the former being $\Theta$-independent and the latter following the same dependence on $\Theta$ as in \citet{marti_structure_2015} and \citet{geng_propagation_2019} (see also \citealt{pavan_jet-environment_2023}): 
\begin{equation}
      B^{r}\!=\!B_\mathrm{ratio} B_0 \,\,\, , \,\, 
      B^{\Phi}(\Theta)=B_0 \frac{2(\Theta/\theta_\mathrm{m})}{1+(\Theta/\theta_\mathrm{m})^2}\quad ,
\end{equation}
where $B_0$, $B_\mathrm{ratio}$, and $\theta_\mathrm{m}$ are constant. 
The jet's magnetic-field configuration was then fully specified by imposing $\theta_\mathrm{m}=0.4 \,\theta_\jet$, and setting the overall field strength through $B_0 = 1.55\times 10^{13}~\G$ and the ratio of radial to azimuthal field via $B_\mathrm{ratio}=0.5$.
\footnote{The specific value of $B_0$ was already employed in one of the models of \citet{pavan_role_2025}, and corresponds to the order-of-magnitude toroidal magnetic field strength at 380\,km radius found by \citet{kalinani_jet-environment_2025} within jets self-consistently launched after black hole (BH) formation.}
Finally, the rest-mass density profile within the jet, $\rho_\jet(\Theta)$, was found by imposing transverse equilibrium and choosing a value for $L_\jet$ (see Appendix~\ref{appendix: transverse_equilibrium}).

In our description, we considered a jet luminosity that decays exponentially in time with characteristic timescale $\tau_L = 0.3~\s$, namely $L_\jet(t) \!=\! L_0 \exp[-(t-t_\jet)/\tau_L]$ with $L_0\!\equiv\!L_\jet(t_\jet)$.
Such a time dependence was enforced by assuming that $\rho_\jet$ and $\vec{B}_\jet$ decrease exponentially with characteristic timescales of $\tau_L$ and $2\tau_L$, respectively.
The chosen exponential decay of $L_\jet$ follows the prescription of \citet{pavan_short_2021, pavan_jet-environment_2023}, where the jet luminosity declines with a characteristic time consistent with the accretion time-scale $\tau_d$ of the BH--disc central engine.
This choice is further motivated by the GRMHD merger simulation of \citet{ciolfi_collimated_2020}, in which the post-merger outflow power exhibits a decline consistent with exponential damping. 
Since the jet breaks out after the engine luminosity has already decreased substantially, the subsequent propagation is primarily determined by the total energy injected before breakout rather than by the detailed temporal profile of the engine luminosity. 
A shorter or longer characteristic timescale would distribute the energy differently, altering the penetrative efficiency of the jet through the ejecta in the first phases. 
As long as the energy is released in a time comparable with the breakout time from the densest inner ejecta, our results are expected to remain qualitatively similar for alternative engine histories \citep[e.g.][]{Gutierrez2025}.

\subsection{Models and numerical methods}
\label{sec: models-methods}

In this study, we considered three models varying the initial jet luminosity and the jet injection time (since merger):
\begin{itemize}
    \item Model A: $t_\jet = 0.185$\,s,\ $L_0 = 10^{51}\,$erg\,s$^{-1}$,
    \item Model B: $t_\jet = 0.385$\,s,\ $L_0 = 10^{51}\,$erg\,s$^{-1}$,
    \item Model C: $t_\jet = 0.385$\,s,\ $L_0 = 5\times10^{51}\,$erg\,s$^{-1}$.
\end{itemize}
These values were selected to sample a broad range of jet--environment interaction regimes, from early breakout to strongly choked propagation, rather than to model any specific event. The adopted one-sided luminosities ($10^{51}$--$5\times10^{51}\,\erg\,\s^{-1}$) exceed those typically inferred for GRB\,170817A; exploring lower-power successful jets is left for future work.
The above $L_0$ values, along with the other fixed parameters (see previous section), correspond to weakly magnetized jets, for which the magnetic contribution to the luminosity is 
\begin{equation}
    \begin{aligned}
        \Sigma_\jet \equiv \dfrac{L_\jet - L_\mathrm{HD}}{L_\jet}
        & \approx 5.51 \% ~ \bigg{(}\frac{L_\jet}{10^{51}~\erg~\s^{-1}}\bigg{)}^{-1}  \quad ,
    \end{aligned}
\end{equation}
where $L_\mathrm{HD}$ denotes the purely hydrodynamic contribution to the jet luminosity.
We note that the inverse scaling $\Sigma_\jet\propto L_\jet^{-1}$ reflects the chosen injection prescription, in which the magnetic-field normalization $B_0$ is kept fixed while the matter contribution to the luminosity is varied; it should not be taken as a physical statement that more luminous jets are more weakly magnetized.

The evolution was carried out using the generalized First ORder CEntred (FORCE) Riemann solver \citep{toro_musta_2006, mattia_comparison_2021}, fifth-order piecewise parabolic reconstruction \citep{mignone_high-order_2014}, and third-order Runge--Kutta time stepping.
We chose a Taub-Matthews equation of state \citep{mignone_equation_2007}, reproducing an ideal gas law with adiabatic index $\gamma\!=\!4/3$ in the ultrarelativistic limit and with $\gamma\!=\!5/3$ in the non-relativistic limit, with a smooth transition at intermediate regimes.
The divergence-free constraint for the magnetic field was enforced using the Hyperbolic Divergence Cleaning \citep{dedner_hyperbolic_2002}, while retaining a cell-centred representation of the primary fluid variables (including the magnetic field) in our simulations.
A vector-force Newtonian field was employed to simulate the gravitational pull from the central compact object, setting its mass to $M_0 \simeq 2.596~\msun$ as in the reference BNS merger simulation.
`Outflow' boundary conditions were imposed at the outer radial boundary and the $\theta$ boundaries, while the conditions were periodic along the $\phi$ direction. The inner radial boundary conditions were user-defined within the chosen half-opening angle, allowing for the jet injection according to the prescription given in the previous subsection (Sect.~\ref{sec: jetprescription}), and set as outflow outside of it. 
During the evolution, we kept track of the fraction of jet material in each computational cell via the scalar $Q_\jet$, passively advected with the fluid. 

\subsection{Extending the evolution in axisymmetry}
\label{sec: axisym}

At the end of the 3D evolution, once the system reaches $t = 0.6$\,s after jet launch, we switched off the jet engine, imposed an outflow condition for the inner radial boundary, and further continued the evolution in 2D and axisymmetry after performing an azimuthal average of the 3D data. 
Before the axisymmetrization step, all fluid quantities were mapped in the tilted coordinate system $(r,\Theta,\Phi)$, in which the polar axis is aligned with the jet injection axis.
During the 3D evolution, the relativistic jet does not necessarily remain aligned with the original injection axis owing to its interaction with the anisotropic environment and the development of fluid instabilities.
A $\Phi$-average taken about the injection axis would smear an off-axis jet into an emitting ring, significantly altering its dynamics during the 2D propagation and modifying the total energy content.
To preserve the energy budget of the original 3D simulation, an appropriate direction must be chosen as the new symmetry axis.
We defined the new symmetry axis as the energy-flux-weighted mean direction $\vec{n}$, computed as the energy-flux-weighted average of the unit radial vectors of all computational cells, $e_\mathrm{tot} \Gamma\beta V$, where $e_\mathrm{tot}$ is the total energy density (including kinetic, thermal, and magnetic energy densities) and $V$ is the volume.
This weighting emphasizes material that is both energetic and relativistic. 
The resulting vector defines the new axis $\vec{n}$, the new polar angle $\Theta''$,  defined as the angle between the original radial direction and $\vec{n}$, and the associated orthonormal basis ($\vec{\Theta}''$, $\vec{\Phi}''$) (see Appendix~\ref{appendix: tilted_jet_axis}). 
Vector fields were properly projected onto the new rotated coordinate basis.
For models A and C, the actual offset with respect to the jet injection axis is very small and thus, for simplicity, we can retain a description based on the same polar axis ($\Theta''=\Theta$), while a redefinition of the main propagation axis is critical for model B (Appendix~\ref{appendix: tilted_jet_axis}).
The azimuthal average (i.e.~along the newly defined $\Phi''$ coordinate) was computed by weighting each quantity with the local energy density.
This choice enhances the imprint of the energetically dominant jet material while suppressing contributions from slower, less relevant regions of the outflow. Moreover, the total energy of the system is better preserved (order $\sim\!1\%$ lower than the original one in 3D) with respect to the non-weighted average (one order of magnitude larger discrepancy).
In this framework, the flow was evolved on a 2D spherical grid while retaining the evolution of $v^{\Phi}$ and $B^{\Phi}$ under periodic boundary conditions. 

Since the relevant part of the jet and cocoon structure is confined within $\Theta\!<\!45^\circ$ from the jet injection axis, we restricted the angular domain to $\Theta \!\in \![0, 45^\circ]$. 
At the same time, we took advantage of the significantly reduced computational cost of 2D simulations to increase the resolution
by a factor of eight while retaining the original cell aspect ratio (i.e.~$\Delta r\!\approx\! r\Delta \Theta$), achieving a finest spacing of $\approx\!0.55$\,km at $r \!=\! 380$\,km.
The corresponding 2D grid was covered by 6144 and 256 points along the radial and $\Theta$ coordinates, respectively.

During the evolution, the inner radial boundary was moved from 380\,km to larger distances at specific times, depending on the considered case. In particular, for model~A, the inner radius was moved to $10^4$\,km at $t - t_\mathrm{j} = 1.8$\,s, while for model~B and~C, it was moved to $10^4$\,km at $t - t_\mathrm{j} = 0.9$\,s, and subsequently to $10^5$\,km at $t - t_\mathrm{j} = 5$\,s. In these steps, the number of grid points is unchanged: the grid remains identical except that points are removed from the inner region and added to the outer region (increasing $r_{\max}$ accordingly).
The inner region that is cut away has no relevance for the dynamics of the jet and cocoon system we are interested in. Moreover, the smallest grid spacing increases, with consequent speedup of the simulated evolution. Finally, the farther away outer boundary allows us to cover longer physical times before the boundary itself is reached by the outflowing material. 

\subsection{Shock front identification and breakout condition}
\label{sec: Shockfront_BO}

To locate the position of the forward shock at each time and angle, we employed the multi-criterion shock-finding procedure described in Appendix~\ref{appendix: shock_finder} (see also 
\citealt{mignone_pluto_2011, gupta_numerical_2021}). 
In order to establish when and where the shock breakout occurs along a given polar angle $\Theta$, we monitored the optical depth at the location of the shock front until the following relativistic shock-breakout condition is satisfied \citep{nakar_electromagnetic_2020}, following the relativistic shock-breakout framework of \citet{nakar_early_2010, nakar_relativistic_2012}:
\begin{equation}
    \tau(r_\sh) \simeq \frac{1}{\beta'_\sh}
    = \frac{1-\beta_\sh\,\beta_\ej}{\beta_\sh - \beta_\ej} \approx \dfrac{\Gamma_\sh^2+\Gamma_\ej^2}{\Gamma_\sh^2-\Gamma_\ej^2} \, ,
\label{bo-condition}
\end{equation}
where $\beta'_\sh$ is the velocity of the shock (in units of $c$) in the local rest frame of the unshocked dynamical ejecta, while $\beta_\sh$ and $\beta_\mathrm{ej}$ are the velocities of the shock and the unshocked dynamical ejecta in the `lab' frame, i.e.~the rest frame of our computational grid and of the observer (with $\beta_\mathrm{sh}>\beta_\mathrm{ej}$).
\begin{figure*}[t]
    \centering
    \includegraphics[width=0.98\linewidth]{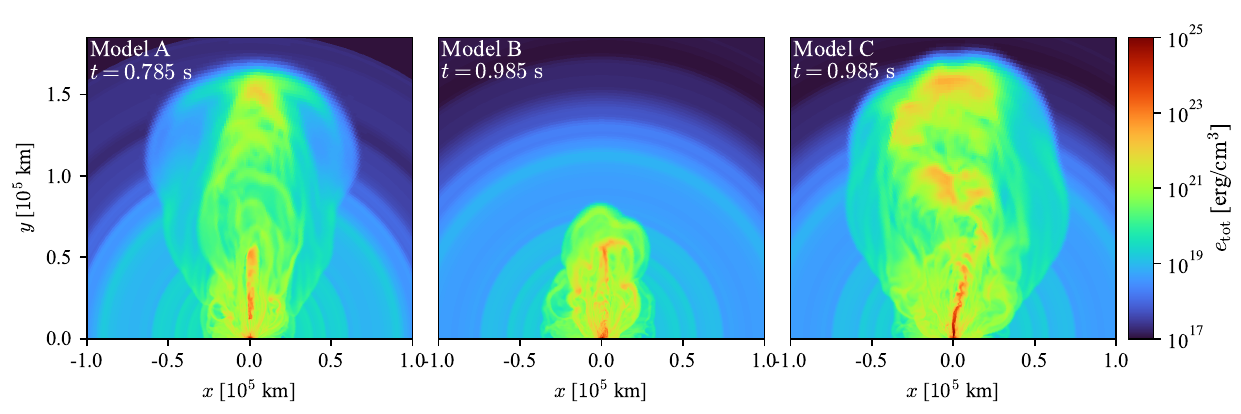}
    \vskip -0.3cm
    \caption{Meridional view of the total energy density $e_\mathrm{tot}$ at $0.6$\,s after the jet injection time for models A, B, and C (left to right). 
    While for models A and C the 2D slice corresponds to the $xy$-plane, for model B we are showing the plane at $\theta = 102.8^\circ$, being the one containing the main jet propagation axis at this time (see text and Appendix~\ref{appendix: tilted_jet_axis} for details).
    The time reported in each panel is the time after merger.} 
    \label{fig: 3panels_06s}
\end{figure*}

At this stage it is sufficient to assume radial photon propagation, i.e.~we integrated the optical depth along the radial direction ahead of the shock; this is adequate to define the breakout condition along each angular direction of the aspherical, laterally expanding jet-cocoon shock, while the more general non-radial photon propagation is treated in the light-curve calculation of Appendix~\ref{appendix: luminosity_curves}.
For a given radial density profile, the optical depth at the shock front is given by \citep{abramowicz_appearance_1991}
\begin{equation}
\label{eq: tau_def1}
    \tau(r_\sh) \;=\; 
    \!\int_{r_\sh}^\infty \kappa \rho(\tilde{r}')\de \tilde{r}' = 
    \!\int^\infty_{r_\sh} \kappa  \rho(\tilde{r})\,\Gamma(\tilde{r}) [1-\beta(\tilde{r})]\,\mathrm{d}\tilde{r} \, ,
\end{equation}
where $\kappa$ is the (frequency independent) opacity, $\tilde{r}'$ and $\tilde{r}$ are the radial length in the local comoving frame of the ejecta and in the lab frame, respectively, and $\Gamma(\tilde{r})$ and $\beta({\tilde{r}})$ are the local Lorentz factor and velocity in units of $c$ as seen in the lab frame. 
As a fiducial value, we took $\kappa \!=\! 0.16~\cm^2/\g$ (see \citealt{nakar_electromagnetic_2020}), while in Sect.~\ref{sec: light_curves_different_opacity} we discuss the effects of varying $\kappa$.
The opacity of neutron-star merger ejecta and its dependence on composition (in particular the lanthanide fraction) have been studied in detail by \citet{kasen_opacities_2013, tanaka_radiative_2013, tanaka_systematic_2020}; our choice lies in the range relevant to the fast, lanthanide-poor outer ejecta probed here.

\section{Evolution results}
\label{sec: evolution_results}

In this section, we present the time evolution of our models A, B, and C (see Sect.~\ref{sec: models-methods}). 
The three models allow us to explore the dependence of jet propagation and breakout on both the launch time (with respect to merger time) and the injected power, while keeping fixed the other incipient jet parameters.
We first describe the initial 3D evolution up to $0.6$\,s after jet launch, which captures the jet interaction with the dense post-merger outflows and the formation of the jet-cocoon structure, and then consider the subsequent axisymmetric evolution extending to $t-t_\jet= 20$\,s, following the long-term propagation of the forward shock into the dilute dynamical ejecta. 

\subsection{3D evolution up to 0.6 seconds}
\label{sec: 3D_evolution}

During its early propagation, the incipient jet has to pierce through the dense post-merger outflows driven by the MNS remnant, transferring part of its energy to the surrounding medium and forming a jet-cocoon structure. 
Upon overcoming the outer radius of the MNS-driven outflows, the jet-cocoon front starts propagating across the much less dense dynamical ejecta, rapidly accelerating and laterally expanding. 

Figure~\ref{fig: 3panels_06s} shows 2D slices of the total energy density $e_\tot$ (including kinetic, thermal, and magnetic energy densities, but not the rest-mass contribution) at $0.6$\,s after jet launch for the three models.  
In model~A (early launch, moderate luminosity), the jet emerges from a comparatively less extended MNS-driven environment, starting earlier its propagation across the outer dynamical ejecta. 
This results in a rather axisymmetric structure, with a narrow energetic jet core surrounded by the less energetic cocoon. 
In contrast, the jet of model~B (later launch time) propagates through a significantly more extended (and massive) MNS-driven environment, losing a larger fraction of its energy while struggling to pierce through. 
As a result, this jet shows clear signs of partial choking, and emerges without a well-defined jet core.

The higher luminosity of model~C partially compensates for the later injection time, producing a more energetic jet-cocoon structure, but less collimated and axisymmetric than model~A.
We note that, while for models~A and C the main jet propagation axis remains essentially aligned with the $y$-axis, for model~B the jet deviates towards a direction about 13 degrees away from the $y$-axis (see Appendix~\ref{appendix: tilted_jet_axis}). For this reason, in Fig.~\ref{fig: 3panels_06s}, the 2D slice shown for model~B corresponds to the plane containing the main jet propagation axis (at $\theta = 102.8^\circ$) instead of the original $xy$-plane.

By $0.6$\,s, the evolution has reached a stage where the global jet-cocoon morphology and angular structure have largely been established. At the same time, the energy injection from the central engine has substantially declined (luminosity a factor $\approx\!7.4$ lower than the initial value).
For the assumed exponential decay, the energy still to be injected after $t-t_\jet=0.6$\,s is $\exp(-2)\approx13.5\%$ of the total. 
This residual energy is released at the base of the outflow, which by this time has decoupled from the jet-cocoon front already propagating into the dilute dynamical ejecta, mostly feeding the slow material at the injection region. Switching off the engine at $0.6$\,s thus leaves the front dynamics and the resulting breakout signal essentially unchanged (see Appendix \ref{appendix: 3D_2D_comparison}).

\subsection{Axisymmetric evolution}
\label{sec: 2D_evolution}

We followed the subsequent evolution via 2D axisymmetric simulations as described in Sect.~\ref{sec: axisym}. 
The transition to axisymmetry preserves the essential features established during the 3D phase --- namely the jet-cocoon morphology and energetics --- while enabling much higher resolution and larger dynamical range at a significantly reduced computational cost.

\begin{figure*}
    \centering
    \includegraphics[width=0.98\linewidth]{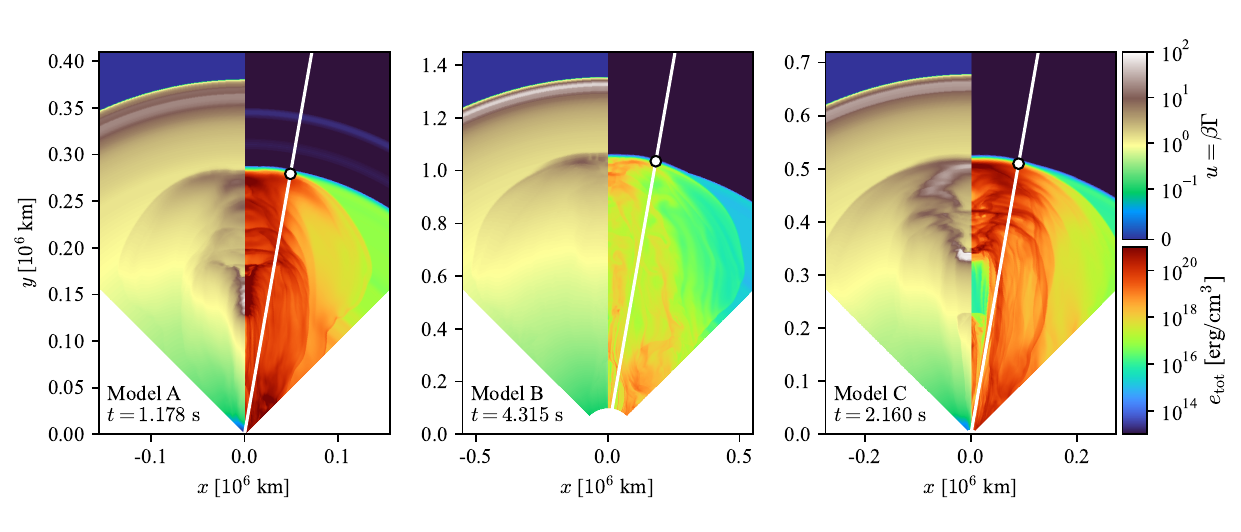}
    \vskip -0.3cm
    \caption{Meridional view of the proper velocity $u=\beta\Gamma$ (left half of the panels) and the total energy density $e_\mathrm{tot}$ (right half of the panels) for the three models A, B, and C at the time when the shock breakout occurred along a radial direction $10^\circ$ away from the jet propagation axis. As in Fig.~\ref{fig: 3panels_06s}, for model B we are showing the plane at $\theta = 102.8^\circ$ instead of the $xy$-plane (see text).   
    The spatial scale in each panel varies depending on the model. The white solid line shows the direction at $10^\circ$, and the white circle marks the breakout radius along that line.} 
    \label{fig: 3panels_breakout_10deg}
\end{figure*}

Figure~\ref{fig: 3panels_breakout_10deg} shows, for each model, the proper velocity $u=\beta\Gamma$ and total energy density $e_\tot$ at the time at which the shock breakout occurs along a radial direction $10^\circ$ away from the jet propagation axis.  
As the flow expands into the outer dynamical ejecta, the forward shock accelerates and gradually approaches a self-similar structure.  
The timing and radius of the breakout differ substantially across the three models.  
Model~A, with early jet injection into a less extended environment, exhibits a much smaller breakout radius $r_\bo$ and earlier breakout time $t_\bo$ compared to model~B, while model~C, whose enhanced luminosity partially compensates for the more extended environment, exhibits intermediate values of $r_\bo$ and $t_\bo$.
We also note that the Lorentz factor at breakout, $\Gamma_\bo$, is smaller than ten for all three models. Beyond the outer radius of the bulk of the dynamical ejecta, at $r\!>\! r_\bo$, there is an extremely low density and energetically negligible portion of material whose front reaches $\Gamma\!>\!10$. As discussed below, this material does not contribute to the optical depth and plays no role in our investigation.
\begin{figure}
    \centering
    \includegraphics[width=0.98\linewidth]{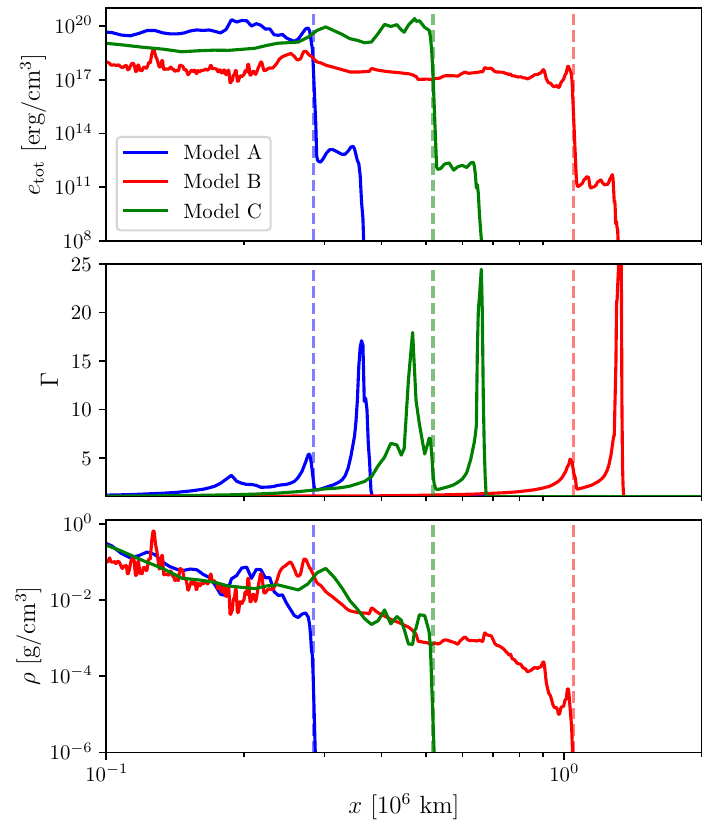}
    \vskip -0.3cm
    \caption{1D profiles of the energy density $e_\mathrm{tot}$ (top), Lorentz factor (middle), and rest-mass density (bottom) along the radial direction $10^\circ$ away from the jet propagation axis, at the corresponding shock breakout time (see Fig.~\ref{fig: 3panels_breakout_10deg}). Different colours refer to different models: A (blue), B (red), and C (green). For each model, a vertical dashed line marks the breakout radius.} 
    \label{fig: 3profiles_breakout_10deg}
\end{figure}

We further note that fast material at large off-axis angles can be affected by the known `plug' artefacts of axisymmetric simulations of jets propagating through a dense medium \citep{zhang_relativistic_2003, gottlieb_structure_2021}. 
In our case, however, the system is axisymmetrized only after the jet breaks out of the densest inner environment and while it is propagating through the much more tenuous outer material. Consequently, the transition from 3D to 2D has only a limited effect on the shock-front evolution and jet propagation (see Fig.~\ref{fig: injection_3D_2D} in Appendix~\ref{appendix: 3D_2D_comparison}), so that the breakout time and the inferred delays are not expected to be significantly affected. Nonetheless, the detailed angular structure at the front of the jet-cocoon system should still be interpreted with caution.

In Figure~\ref{fig: 3profiles_breakout_10deg}, we present the radial profiles of total energy density, Lorentz factor, and rest-mass density along the direction $10^\circ$ away from the jet propagation axis at the corresponding breakout times (i.e.~the same times indicated in Fig.~\ref{fig: 3panels_breakout_10deg}). 
The shock front radius $r_\bo$ is marked by a vertical dashed line.
All models exhibit a sharp, well-defined shock front, as seen in the energy density profile.
For model~B, the rest-mass density drop prior to the shock front is much more gradual and the peak of the velocity at the front is significantly lower. These features reflect the fact that, in this case, the jet is significantly choked (see also Fig.~\ref{fig: 3panels_breakout_10deg}). 

For all three models, looking at distances larger than $r_\bo$, the profiles clearly show that the energy content of the outer material is negligible, despite the presence of a rather high peak in Lorentz factor. 
Moreover, due to the extremely low densities, the contribution of this material to the optical depth is also negligible.

\section{GW-EM delay and shock-breakout luminosity}
\label{sec: gw_em_delay}

In this section, we compute the time interval between the detection of the GW signal (emitted at the merger time) and the arrival of the first photons produced during the shock breakout of the jet-cocoon system, depending on the observing angle. Moreover, we compute the isotropic-equivalent bolometric luminosity of the signal, under the assumption that the thermal energy of the shocked material is entirely converted into radiation.
\begin{figure*}
    \centering
    \includegraphics[width=1\linewidth]{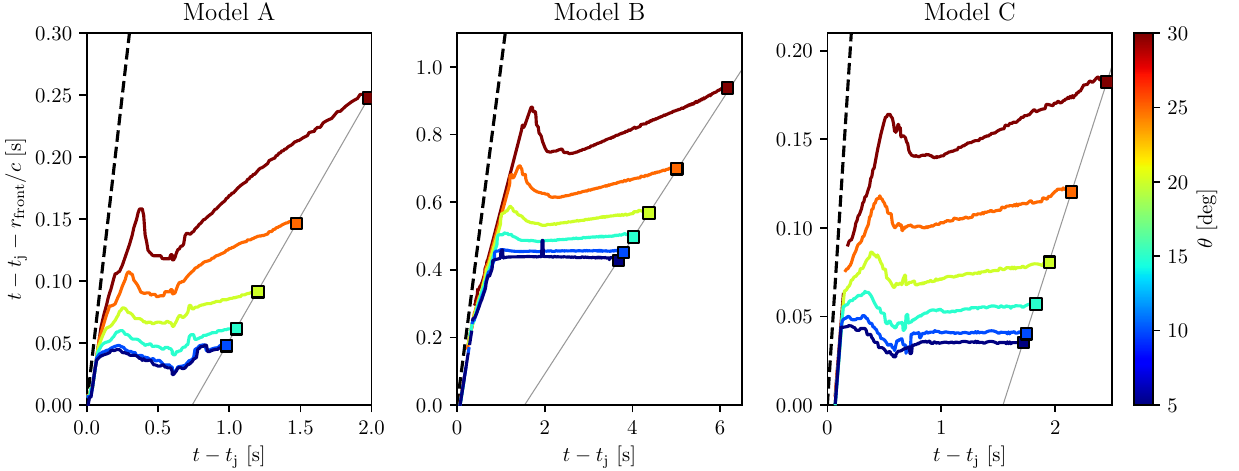}
    \vskip -0.3cm
    \caption{Delay accumulated over time by the jet-cocoon shock front (at $r=r_\mathrm{front}$) with respect to the GW signal, excluding the jet launch time $t_\jet$, along different polar angles (colour-coded) and for the different models. 
    An opacity of $\kappa = 0.16\,\cm^2/\g$ is assumed.
    The coloured squares mark the shock breakout. The black dashed line in each panel represents the delay that would be accumulated in the limit of a shock front advancing with zero velocity (i.e.~at rest, with constant $r_\mathrm{front}=380$\,km).
    The thin grey line is the outer radius of the bulk of the dynamical ejecta, expanding self-similarly at about $0.8\,c$.} 
    \label{fig: GWEM_delay_accumulation}
\end{figure*}
\setlength{\tabcolsep}{5pt}
\begin{table}
    \centering
    \begin{tabular}{c|cccccc}
        Model & $t_\jet$ [s] & $L_{0, 51}$ & $r_{\bo,10}$ & $t_\bo$ [s] & $\Gamma_\bo$ & $\Delta t_\delay$ [s]\\
        \hline
        A & 0.185 & 1 & 2.8 & 1.17 & 3.7 & 0.24 \\
        B & 0.385 & 1 & 10.5 & 4.32 & 3.6 & 0.81 \\
        C & 0.385 & 5 & 5.1 & 2.11 & 5.4 & 0.42 \\
    \end{tabular}
    \caption{Shock breakout properties at $5^\circ$ from the jet propagation axis, for models A, B, and C, with a different jet launch time $t_\jet$ (since merger time) and initial luminosity $L_{0}$ (expressed in units of $10^{51}~\erg/\s$).
    For each model, we report breakout radius $r_\bo$ (expressed in units of $10^{10}~\cm$), breakout time since merger $t_\bo$, Lorentz factor at breakout $\Gamma_\bo$, and GW-EM delay time $\Delta t_\delay$. The assumed opacity is $\kappa=0.16~\cm^2/\g$.}
    \label{tab: table_breakout_5deg}
\end{table}
\setlength{\tabcolsep}{6pt}

\subsection{Time evolution and angular dependence of the delay}
\label{sec: delayresults}

We first consider the delay along a fixed angular direction $\Theta$, given by $t_\bo-r_\bo/c$, where $t_\bo$ is the time since merger at which breakout occurs and thus photons are free to propagate at the speed of light towards the observer. 
In Table~\ref{tab: table_breakout_5deg}, we summarize the results for our three models at $5^\circ$ from the jet propagation axis, in terms of $r_\bo$, $t_\bo$, $\Gamma_\bo$ and the delay itself ($\Delta t_\delay$).
We note that the given delays include the time interval between merger and jet launch. Once this time interval is subtracted, the delay due to the jet propagation until breakout becomes about 60, 660, and 40\,ms, respectively. 
We thus infer that, for models A and C, in which either a less massive and extended environment or a factor of five higher luminosity significantly enhances jet propagation, the total GW-EM delay is largely determined by the time interval between merger and jet launch, which accounts for about 77\% and 92\% of the total delay, respectively. 
Conversely, for model B, in which jet propagation is less efficient, this interval accounts for only 37\% of the total delay, with the propagation itself providing the largest, though comparable, contribution.

In Figure~\ref{fig: GWEM_delay_accumulation}, we illustrate how the GW-EM delay accumulates over time until breakout, for several polar angles (from $5^\circ$ to $30^\circ$, at steps of $5^\circ$) and for the different models. Each profile ends at the shock breakout time, represented by a square marker.
We note that the delay, in the early phase, is not necessarily monotonically increasing. This is due to the non-radial component of the shock front propagation: along a given polar angle, when a more advanced and laterally expanding shock front becomes detectable, the reference radius for computing the delay can move ahead faster than the speed of light (while no fluid element is violating causality). 

The thin grey lines in Fig.~\ref{fig: GWEM_delay_accumulation} trace the outer edge of the bulk of the dynamical ejecta, which are isotropic and move self-similarly with constant velocity $v_\mathrm{ej} \!\simeq\! 0.8\,c$.
The corresponding radius is 
\begin{equation}
\label{eq: r_de_last}
    R_\mathrm{ej}(t) = 9.24 \times 10^9~\cm \left( \dfrac{t}{0.385~\s}\right) \quad ,
\end{equation}
where $t$ is the time since merger. 
Shock breakout along a given direction occurs when the jet-cocoon shock front reaches such an edge. 

After a first early phase (at $\gtrsim\!1$\,s since jet launch), the jet-cocoon shock front can be approximated as an expanding ellipsoid, with semi-major axis $r_\parallel(t) = r_0 + v_\parallel (t-t_\jet)$ and semi-minor axis $r_\perp(t) = v_\perp (t-t_\jet)$, with $r_0\!=\!380$\,km and where $v_\parallel$ and $v_\perp$ are the expansion velocities along the jet injection axis and along the direction orthogonal to it, respectively. The corresponding ellipsoidal radius is 
\begin{equation}
    r_\mathrm{ellipsoid}(t,\Theta) = \left[ \dfrac{\cos^2\Theta}{r_\parallel^2(t)} + \dfrac{\sin^2\Theta}{r_\perp^2(t)} \right]^{-1/2} \, .
\end{equation}
The condition $r_\mathrm{ellipsoid}(t,\Theta) = R_\mathrm{ej}(t)$ determines a fourth-order polynomial in $t$ whose real positive root is the shock breakout time $t_\bo(\Theta)$.
The on-axis breakout is easily found imposing $\Theta=0$ and solving for $t$:
\begin{equation}
   t_\bo(\Theta=0) = \dfrac{v_\parallel t_\jet- r_0}{v_\parallel-v_\mathrm{ej}} \simeq \dfrac{v_\parallel t_\jet}{v_\parallel - v_\mathrm{ej}} \, .
\end{equation}
Comparing the on-axis breakout time with the simulation results for each model, we can determine $v_\parallel$, while $v_\perp$ is left as a free parameter. 
The values for $v_\parallel$ are found to be $(0.963,0.886,0.983)\,c$ for models A, B, and C, respectively.
Then, fixing $v_\perp$ to $(0.710,0.764,0.801)\,c$ for the three models, we are able to match all breakout times up to $30^\circ$ within a $5\%$ error. 
This shows that a simple analytic description, without explicitly modelling the complex jet dynamics and momentum balance equation for jets embedded in an expanding medium \citep{hamidani_jet_2021}, can capture the main features of the jet-cocoon shock expansion at times $\gtrsim\!1$\,s since jet launch. 

\subsection{Relativistic beaming}
\label{sec:beaming}

At breakout, the relativistic beaming confines the radiation emitted from the shock front at $\Theta$ within a half-opening angle $\theta_\beam$, where $\sin\theta_\beam \!=\! 1/\Gamma_\bo(\Theta)$ and $\cos\theta_\beam \!=\! \beta_\bo(\Theta)$.
The first shock breakout photons received by an observer at $\Theta\!=\!\theta_\obs$ may originate from a portion of the shock front moving along a different polar angle, and characterized by a smaller $t_\bo$, as long as the observer direction is within the beaming angle of the emitting region. 
Since $t_\bo$ is monotonically increasing with $\Theta$, for a given $\theta_\obs$ we computed the delay of the first photons emitted from the shock front at a $\Theta$ such that $\Theta+\theta_\beam(\Theta)=\theta_\obs$, and compare with the delay of the photons emitted at $\Theta\!=\!\theta_\obs$. 
The actual delay observed will be the lesser of the two.

\begin{figure}
    \centering
    \includegraphics[width=1\linewidth]{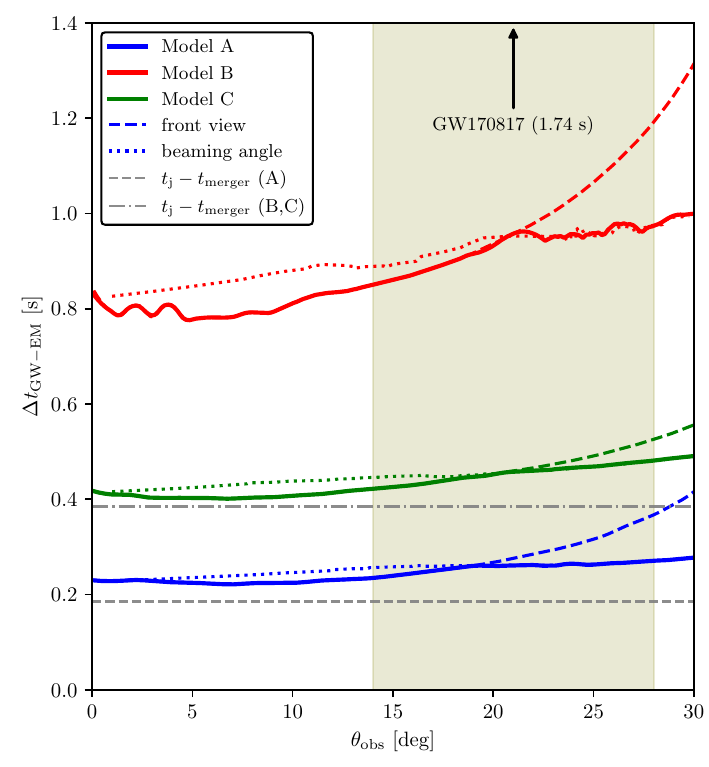}
    \vskip -0.3cm
    \caption{Time delay between the GW and the shock breakout signal ($\Delta t_\delay$) as a function of the observing angle $\theta_\mathrm{obs}$, for our three models (represented in different colours) and assuming an opacity of $\kappa = 0.16~\cm^2/\g$. 
    The dashed and dotted lines show, respectively, the delays computed for photons emitted from a shock region at either (i) a polar angle equal to $\theta_\mathrm{obs}$ or (ii) the smallest polar angle for which the relativistic beaming cone still includes the observer direction.  
    The minimum of the two, indicated with a thicker continuous line, is the actual delay (see text for details).
    The shaded region marks the observationally inferred viewing-angle range for GW170817. The horizontal grey lines indicate the jet launching time relative to the merger time for each model, as reported in the legend.} 
    \label{fig: GWEM_delay}
\end{figure}

In Figure~\ref{fig: GWEM_delay}, for each model, we report both delays (front-view and beaming-corrected) depending on $\theta_\obs$, and mark with a thicker continuous line minimum of the two (representing the actually observed delay). 
For $\theta_\obs\gtrsim20^\circ$, photons emitted closer to the jet axis arrive earlier and this effect results in a significantly weaker dependence of the delay itself on $\theta_\obs$.

Overall, $\Delta t_\delay$ is found to depend rather poorly on the observer direction: going from $0^\circ$ to $30^\circ$, we find variations of $\approx\!0.3$\,s for model B and $<\!0.1$\,s for models A and C.
This property, at least for the specific model set explored here, makes $\Delta t_\delay$ a powerful diagnostic to distinguish among models. 
In general, for each model having such a poor angle dependence, when comparing to an observation with precise delay determination but a wide range of possible viewing angles (as in the case of GW170817), the model itself can be confidently excluded unless the narrow range of possible delays is consistent with the observed one. 

Taking the range $\theta_\mathrm{obs} \in [14^\circ, 28^\circ]$ as representative for GW170817 \citep{mooley_superluminal_2018, mooley_optical_2022}, Table~\ref{tab: model_range_GW170817} shows for our three models the corresponding range of possible delays, to be compared with the $\approx\!1.74$\,s of GW170817 \citep{LVC-GRB}.
In all cases, variations are within 15\% and compatibility with the 2017 event can be excluded with full confidence. In terms of delay only, model B is significantly closer to 1.74\,s, although still inconsistent.

\subsection{Emitted luminosity}
\label{sec: emitted_luminosity}

We computed the EM emission associated with shock breakout using the procedure described in detail in Appendix~\ref{appendix: luminosity_curves}.
Here we summarize the essential elements of such a procedure.
For a given observing angle $\theta_\obs$, the shock-breakout time $t_\bo$ is defined as the time at which the shock reaches the photospheric or emission radius, determined by the condition $\tau \!\simeq\! 1/\beta'_\sh$ (see Eq.~\ref{bo-condition}). 
After this time, photons initially trapped in the shocked material start diffusing out. The emission radius, which changes in time according to the same condition $\tau \!\simeq\! 1/\beta'_\sh$, 
recedes in time (moving backwards in the frame comoving with the shock front) as radiation escapes from progressively deeper regions of the flow. Assuming that the signal is powered by the thermal energy contained within the emitting layer, the contribution from each portion of the emitting layer in a given timestep will correspond to the increase of thermal energy outside the emission radius. 

To account for non-radial contributions, at each timestep we divided the emitting layer in portions that, projected on a plane orthogonal to the observer's direction, correspond to $(2N+1)^2$ equal squares of side length $\lambda=R/N$, where $R$ is the time-evolving radial extension of the emitting layer and $N=50$. For each portion, we computed the optical depth by integrating along the observer's direction, establishing the volume containing the thermal energy that will be released in that timestep (see Fig.~\ref{fig: non_radial_contribution}). Moreover, we accounted for relativistic Doppler shifts towards the observer and for the different light travel times from each portion. As final result, we obtained the observed isotropic-equivalent bolometric luminosity $\mathcal{L}_\iso$ as function of the observing time $T_\obs$, 
for the chosen $\theta_\obs$.
\begin{table}
    \centering
    \begin{tabular}{c|ccc}
        Model & $t_\jet$ [s] & $L_0$ [$10^{51}~$erg/s]  & \begin{tabular}[c]{@{}c@{}}
        $\Delta t_{\delay}$ [s] \\
        for $\theta_\mathrm{obs,GW170817}$
    \end{tabular} \\
        \hline
        A & 0.185 & 1 & 0.24 -- 0.28 \\
        B & 0.385 & 1 & 0.84 -- 0.94 \\
        C & 0.385 & 5 & 0.44 -- 0.49 \\
    \end{tabular}
    \caption{Time delays between the GW and the shock breakout signal ($\Delta t_\delay$) corresponding to the range of viewing angles $\theta_\mathrm{obs} \in [14^\circ, 28^\circ]$, representative of GW170817.}
    \label{tab: model_range_GW170817}
\end{table}

\begin{figure}
    \centering
    \includegraphics[width=1\linewidth]{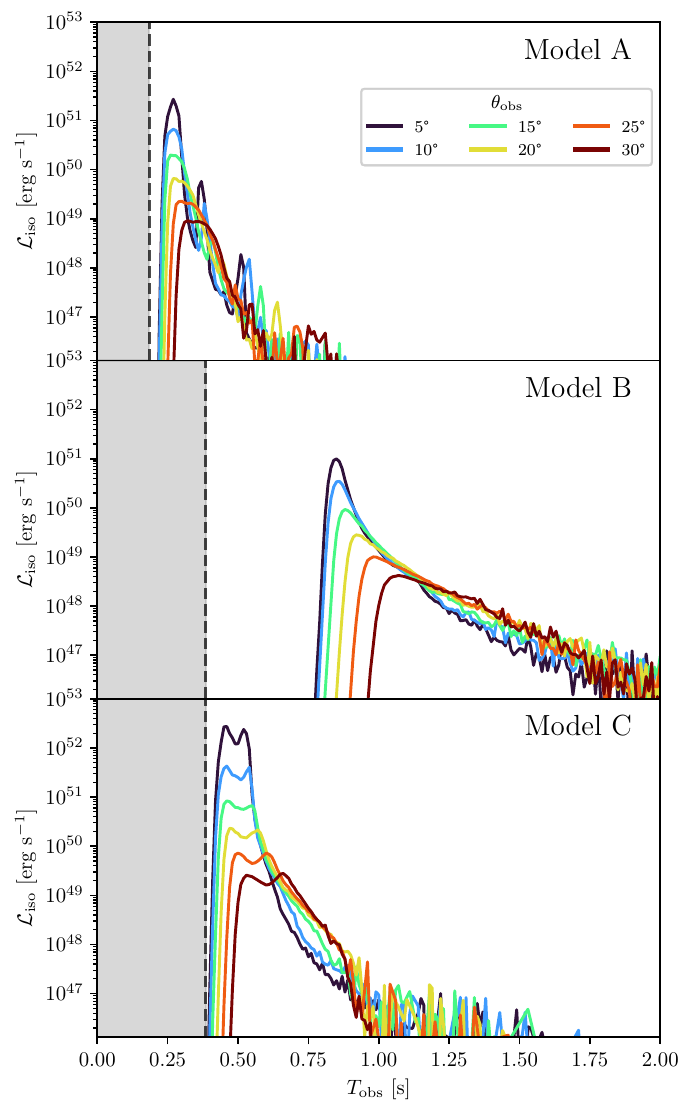}
    \vskip -0.3cm
    \caption{Isotropic-equivalent bolometric luminosity of the shock breakout emission for different models (top to bottom) and viewing angles (colour-coded), as a function of the observer time $T_\obs$ (where $T_\obs=0$ corresponds to the arrival time of the GW signal). 
    The grey area extending up to the vertical black dashed line corresponds to the time interval between merger and jet launch.} 
    \label{fig: light_curves}
\end{figure}

The resulting $\mathcal{L}_\iso$ for models~A, B, and C are shown in Figure~\ref{fig: light_curves} for several viewing angles $\theta_\obs$ (from $5^\circ$ to 30$^\circ$ from the jet propagation axis, at steps of $5^\circ$).
The grey shaded region marks the time interval from merger to jet launch, during which no emission is expected. 
In all models, the luminosity rapidly rises following shock breakout, and after reaching its peak, it decays on a sub-second timescale.  
Model~A and C produce fast-rising emission early after jet launch, while model~B yields a significantly delayed and broader signal, consistent with its later breakout. 

The onset time of the emission with respect to the GW arrival time can be established in a model-independent way by computing $T_{5\%}$, i.e.~the time at which the 
integrated luminosity corresponds to 5\% of the total emitted energy.
The values of $T_{5\%}$ for the different models and viewing angles are reported in Table~\ref{tab: T5}.
We remark that such onset times are fully consistent with the GW-EM delays ($\Delta t_\delay$) previously computed based on the shock breakout condition only (see Sect.~\ref{sec: delayresults} and Sect.~\ref{sec:beaming}).

\begin{figure}
    \centering
    \includegraphics[width=1\linewidth]{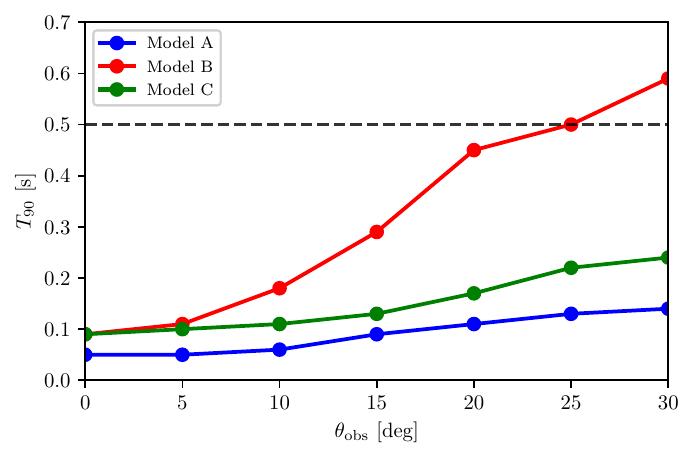}
    \vskip -0.3cm
    \caption{Fluence time $T_{90}$ for the different models as a function of the viewing angle. The horizontal black dashed line represents the approximate duration of the main peak of GRB\,170817A.} 
    \label{fig: T90}
\end{figure}

In Figure~\ref{fig: T90}, we show instead the time interval $T_{90}$, going from $T_{5\%}$ to $T_{95\%}$, depending on the model and the viewing angle.
As expected, results for models A and C are rather similar, with a $T_{90}$ ranging from about 0.05 to 0.2\,s, while for model B we obtain significantly larger values in the range $\approx\!0.1-0.6$\,s.
These values are of the same order as the $\approx\!0.5$\,s of the main peak of GRB\,170817A \citep{goldstein_ordinary_2017}, but we caution that this result is only qualitative. We are directly comparing bolometric and Fermi/GBM-band durations, without taking into account the potentially significant effects of spectral evolution and detector bandpass.
In addition, we note that the duration of the breakout signal is in order-of-magnitude agreement with the dynamical timescale $r_\bo/(\Gamma^2_\bo \beta_\bo c)$.

In terms of peak luminosities, besides the strong dependence on the viewing angle, the outcome changes by orders of magnitude among the three models (Figure~\ref{fig: light_curves}).
At $\theta_\obs\!=\!15^\circ$, for instance, model B reaches $\sim 10^{50}$\,erg/s, model A $\sim 2\times10^{50}$\,erg/s, and model C $\sim 10^{51}$\,erg/s.

The computed $\mathcal{L}_\iso$ is bolometric and obtained under the assumption that all the available thermal energy is converted into radiation.
In order to relate to a signal such as GRB\,170817A, whose inferred isotropic-equivalent peak luminosity in the high-energy band (1\,keV--10\,MeV) is $\simeq 1.6 \times 10^{47}$\,erg/s \citep{goldstein_ordinary_2017, zhang_peculiar_2018}, one would need to estimate the fraction of $\mathcal{L}_\iso$ that would contribute to the observed signal in the same high-energy band, i.e. the factor $f_\gamma\equiv \mathcal{L}_{\iso,\gamma}/\mathcal{L}_\iso$ whose precise value depends on the (here unmodelled) shock-breakout spectrum.
Restricting to $\theta_\obs$ between $15^\circ$ and $25^\circ$, i.e.~roughly the range of possible viewing angles consistent with GRB\,170817A, the values of $f_\gamma$ that would reconcile our three models with such a signal are: $\sim 10^{-3} - 10^{-2}$ for models~A and B, and $\sim 2\times10^{-4} - 2\times10^{-3}$ for model C.

\begin{table}
    \centering
    \begin{tabular}{c|c|c|c}
    \multirow{2}{*}{$\theta_{\rm obs}$ [deg]} 
    & \multicolumn{3}{c}{Onset time ($T_{5\%}$) [s]} \\
    \cline{2-4}
    & Model A & Model B & Model C \\
    \hline
        5  & 0.24 & 0.83 & 0.44 \\
        10 & 0.25 & 0.83 & 0.44 \\
        15 & 0.26 & 0.86 & 0.45 \\
        20 & 0.27 & 0.90 & 0.47 \\
        25 & 0.28 & 0.96 & 0.49 \\
        30 & 0.30 & 1.03 & 0.52 \\
    \end{tabular}
    \caption{Onset time of the shock breakout signal with respect to the GW arrival time, for the different models and viewing angles, computed as $T_{5\%}$ (i.e. the time at which the emitted energy reached $5\%$ of the total).}
    \label{tab: T5}
\end{table}

\subsection{Light curves with different opacities}
\label{sec: light_curves_different_opacity}

\begin{figure}
    \centering
    \includegraphics[width=1\linewidth]{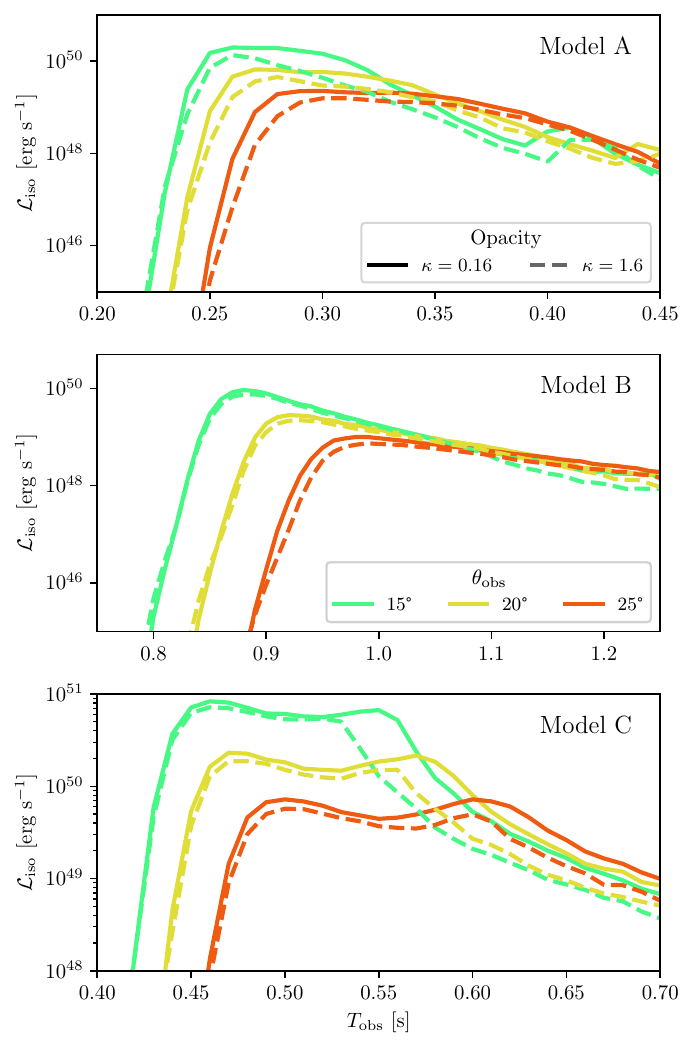}
    \caption{Same as Fig.~\ref{fig: light_curves}, but now considering two different opacity values: $\kappa=0.16$\,cm$^2$/g (continuous lines) and $1.6$\,cm$^2$/g (dotted lines). In this case, we only show three viewing angles, i.e.~15, 20, and 25 degrees, maintaining the same colours employed in Fig.~\ref{fig: light_curves}.
    }
    \label{fig: different_opacities}
\end{figure}

Figure~\ref{fig: different_opacities} shows the isotropic-equivalent bolometric luminosity of the shock breakout emission for different models (top to bottom) and viewing angles (colour-coded), as a function of the observer time $T_\obs$ (with the arrival time of the GW signal set to $T_\obs=0$), where we consider two different constant values for the opacity: $\kappa=0.16$\,cm$^2$/g (fiducial value) and ten times larger $\kappa=1.6$\,cm$^2$/g.
We find that a factor of ten increase in opacity does not sensibly alter the onset time of the signal, altering it by $<\!1.5\%$ for all models. 
This indicates that the dependence of the GW-EM delay on the opacity is very weak, making such a diagnostic even more effective in comparing models with a specific observation. 

Considering the overall light curves, the increased value of $\kappa$ has some limited effects.
In particular, the peak luminosity is lower by at most a factor of two and the duration $T_{90}$ is smaller by less than 30\%.

\section{Summary and conclusions}
\label{sec: conclusions}

In this work, we investigated the time delay between the peak GW signal of a BNS merger and the onset of shock-breakout EM emission associated with a relativistic jet launched after merger and piercing through the surrounding medium.
To this end, we performed RMHD simulations of jet propagation in a realistic post-merger environment, directly imported from the outcome of a previous GRMHD BNS merger simulation. 
Such a propagation environment includes the outer dynamical ejecta component, which plays a crucial role in regulating the shock-breakout condition and thus the corresponding breakout time and radius along any angular direction.
Since this key component is suppressed by the high numerical density floor in the GRMHD simulation, we reconstructed it from the original matter outflow at 300\,km distance, where no suppression has occurred yet. For simplicity, we reconstructed the 1D dynamical ejecta profile along the orbital (or jet-injection) axis, and then assumed a spherically symmetric distribution. The evolution of the system was followed in 3D up to 0.6\,s (since jet launch) and then continued in 2D up to several seconds.
\footnote{Moving to a 2D evolution allowed us to increase resolution by a factor of eight, thus maintaining a small enough grid spacing up to the very large distances we needed to reach. This made our investigation feasible with our current computational capabilities. Nonetheless, it represents a caveat that should be taken into account.}

By varying the jet launching time (since merger) and the injected luminosity, we considered three representative models (A, B, and C) that span a range of jet-environment interaction regimes: (i) early and relatively clean breakout, (ii) slower propagation through a more extended and massive surrounding medium, (iii) relatively fast propagation through the more extended medium enabled by a higher jet power. 
These models allowed us to explore how different physical conditions affect both the timing and early luminosity of the shock-breakout emission.

For each model, we tracked the forward shock in front of the jet-cocoon system up to the photosphere, defined by the condition $\tau \!\simeq\! 1/\beta'_{\rm sh}$, where $\beta'_{\rm sh}$ is the shock velocity in units of $c$ relative to the unshocked dynamical ejecta.
This enabled us to determine self-consistently the onset time of the shock-breakout emission and to compute the corresponding observed GW-EM delay.
The calculation explicitly accounts for geometric effects, light-travel-time differences, and relativistic beaming, and was carried out for a range of observer's viewing angles, adopting a fiducial (frequency independent) opacity of $\kappa=0.16$\,cm$^2$/g.

A key result of our investigation is the weak dependence, found for the considered set of models, of the GW-EM delay on the viewing angle.
Between the on-axis view and $\theta_{\rm obs} \!=\! 30^\circ$, it varies only by about 20\% for models A and C, and 30\% for model B. 
Such a weak dependence makes this time interval, at least for the explored models, a robust diagnostic to exclude specific jet-launch scenarios independently of the poorly constrained observer orientation. The magnitude of the delay itself, and hence which models are excluded, remains configuration-dependent.

Comparing our limited set of models with the observed GW-EM delay of GW170817, we found that our jets are too powerful and/or launched too early.  
Among the cases explored, model B provides the largest time delay, featuring the most extended interaction between the jet and the surrounding ejecta and yielding a delay in the range $\approx0.8 - 1$\,s for $\theta_{\rm obs}$ in the expected viewing range of GW170817.
Nonetheless, the jet is substantially choked and likely inconsistent with other observational signatures (e.g. the jet afterglow signal).
These indications will be a reference to start a more systematic exploration of the parameter space, searching for models compatible with GW170817 and, at the same time, investigating in more detail how the different properties of the incipient jet and/or the post-merger environment affect the GW-EM delay.

As a further step, we extended our analysis to the EM emission associated with shock breakout. 
In particular, we computed the isotropic-equivalent bolometric luminosity resulting from the thermal energy released over time by the shocked material, depending on the viewing angle. In this calculation, we took into account relativistic Doppler shifts, and light-travel-time effects.  
Since the late-time evolution was followed under the assumption of axisymmetry, we reconstructed the full 3D system in terms of 2D slices through a tomographic procedure. This allows us to take into account non-radial photon propagation from each portion of the whole emitting layer.

To characterize the onset of the emission in a quantitative and model-independent way, we introduced the time $T_{5\%}$ at which the time-integrated luminosity reaches 5\% of the total emitted energy. 
We found that the corresponding onset times closely track the GW-EM delays obtained from the previous shock-breakout analysis, providing an internal consistency check of our results. 
As found for the delay, the dependence of $T_{5\%}$ on the viewing angle is weak, while significant differences arise among the different jet-launch scenarios.

Although the emission mechanism responsible for the observed $\gamma$-ray signal of GRB~170817A remains uncertain, we performed a first qualitative comparison between our light curves and the observed prompt emission. 
Since our luminosity is bolometric and assumes full conversion of thermal energy into radiation, it is expected to exceed that in the reference high-energy band (i.e.~1\,keV--10\,MeV).
The peak luminosity of models A and B could be reconciled with that of GRB\,170817A with a radiative efficiency factor between $\sim 10^{-3}$ and $10^{-2}$, whereas the higher luminosities of model C appear more difficult to accommodate.

Another comparison was made in terms of the observed duration of the prompt emission. The measured $T_{90}$ of GRB\,170817A is approximately $2$\,s for the full burst, with a main peak lasting about $0.5$\,s, as measured in the Fermi/GBM energy band.
Depending on the model and the viewing angle, the $T_{90}$ of our bolometric luminosities range from about 0.05 to $0.6$\,s, with model B having the largest durations and thus resulting, also in this case, the closest one to the main peak of GRB\,170817A.
We remark, however, that this comparison is only qualitative and should be taken with caution: our $T_{90}$ is bolometric, whereas the observed one is measured in a finite detector bandpass, and spectral evolution together with the bandpass can significantly affect the inferred duration.

Finally, we computed the shock breakout luminosity with a ten times larger opacity ($\kappa=1.6$\,cm$^2$/g).
We found that the onset time of the emission is poorly affected ($<1.5\%$ difference), implying a very weak dependence of the GW-EM delay on the opacity. This further reinforces the robustness of this observable as a tool to constrain models.
Looking at peak luminosity and signal duration, the increased opacity can have non-negligible effects.
In particular, depending on the model and the viewing angle, peak luminosities are reduced by at most a factor of two, while $T_{90}$ is smaller by no more than 30\%. 
In future work, in addition to performing a wide parameter exploration, we aim at improving our modelling of the shock-breakout emission, thus enabling more reliable comparisons with the GRB\,170817A signal in terms of luminosities and spectra. 

\begin{acknowledgements}
We thank the referee for the useful comments that helped to sensibly improve the quality of the work.
The authors acknowledge support from the European Union under NextGenerationEU via the PRIN 2022 Project `EMERGE', Prot.~n.~2022KX2Z3B (CUP C53D23001150006). RC acknowledges further support from the INAF Theory Grant 2023 `AfterJet', Ob.Fu.~1.05.23.06.02 (CUP C93C23006800005).
AP is partially supported by the INAF Mini Grant 2024 `End2End', Ob.Fu.~1.05.24.07.04 (CUP C93C24008030001).
Simulations were performed on the Discoverer (Sofia Tech Park, Bulgaria) and Leonardo (CINECA, Italy) HPC clusters.
We acknowledge EuroHPC Joint Undertaking for awarding us access to Discoverer via the Regular Access allocation EHPC-REG-2025R02-007, and CINECA for the availability of high performance computing resources and support on Leonardo through an award under the ISCRA initiative (Grant IsB32 MerJet).
\end{acknowledgements}

\bibliographystyle{bibtex/aa} 
\bibliography{bibtex/main} 

\begin{appendix}
\nolinenumbers
\counterwithin{equation}{section}
\counterwithin{figure}{section}

\section{Tilted reference frame}

\subsection{Initial conditions tilt}
\label{appendix: tilted_reference_frame}
In our 3D simulations, jets are injected along a direction that is orthogonal to the polar axis of the spherical coordinates, i.e.~along $\theta,\phi=\pi/2$. 
However, to continue the simulation in 2D under the assumption of axisymmetry, we first needed to convert all our data (including vector components) to a spherical coordinate system ($r,\Theta,\Phi$) with polar axis aligned with the jet injection axis.
Applying the necessary rotation, $\Theta$ and $\Phi$ are related to $\theta$ and $\phi$ via the following expressions:
\begin{equation}
\label{eq: theta_phi_prime}
    \begin{aligned}
        &\Theta = \arccos(\sin\theta\sin\phi) \quad ,\\
        &\Phi =  \pi -\arctan\left( \dfrac{\cot\theta}{ \cos\phi}\right) \quad .
    \end{aligned}
\end{equation} 

For a generic vector in 3D spherical coordinates $\vec{A} = (A_r, A_\theta, A_\phi)$, the relations are 
\begin{equation}
\label{eq: spherical_fields_rotated}
    A_\theta = \dfrac{\cos\phi}{\sin\Theta} A_{\Phi} \,\,,\quad 
    A_\phi = -\dfrac{\cos\theta\sin\phi}{\sin\Theta} A_{\Phi} \quad .
\end{equation}

\subsection{Axisymmetrization around tilted jet axis}
\label{appendix: tilted_jet_axis}

To identify the main direction along which the jet propagates in the 3D simulation, we computed a weighted average of all cells as follows. 
We consider the unit vector on a sphere
\begin{equation}
    \vers{n}^{\mathrm{sph}}_{ijk} =\big(\sin\theta_{ij}\cos\phi_{ik},\ \sin\theta_{ij}\sin\phi_{ik},\ \cos\theta_{ij}\big) \, ,
\end{equation}
where $i, j$ and $k$ are the radial, polar and azimuthal indices, respectively. 
For each cell in the domain, we defined a weight based on the local energy density flux, $w_{ijk} = [e_\mathrm{tot}\Gamma\beta]_{ijk} \de V_{ijk}$, such that the mean direction $\vec{n}$ is 
\begin{equation}
\label{eq: n_jet}
    \vec{n} = \frac{\sum_{ijk} w_{ijk} \vers{n}^{\mathrm{sph}}_{ijk}}{\sum_{ijk} w_{ijk}} \, .
\end{equation}
The centroid position $(\Theta_\mathrm{centr}, \Phi_\mathrm{centr})$ is then given by 
\begin{equation}
\label{eq: centroid_coord}
    \Theta_\mathrm{centr} = \arccos(\vec{n}\cdot\vers{y}) \qquad \Phi_\mathrm{centr} = \arctan[(\vec{n}\cdot\vers{z})/(\vec{n}\cdot\vers{x})] \, .
\end{equation}
\begin{figure}
    \centering
    \includegraphics[width=0.76\linewidth]{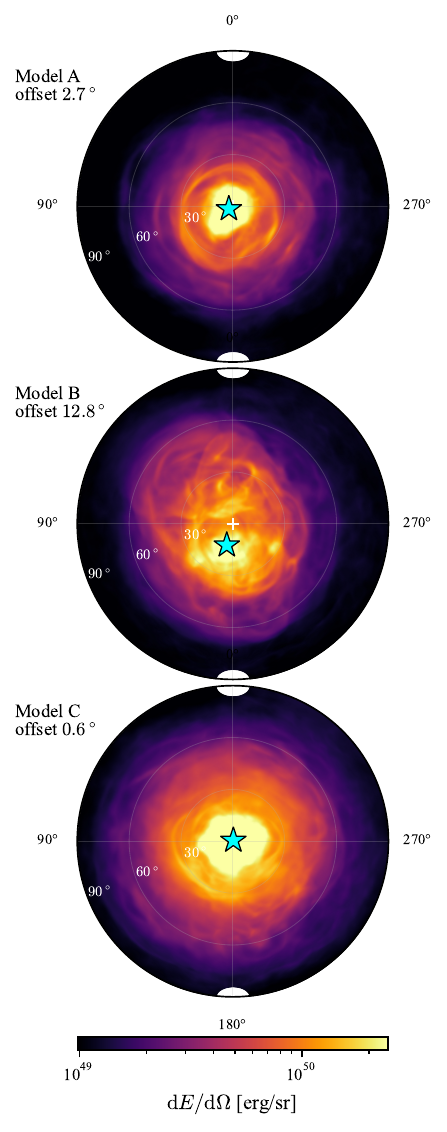}
    \caption{Polar plots in the ($\Theta,\Phi$) coordinate system showing the total energy per unit solid angle for the three models at 0.6 s after the jet launch. The blue star in each panel indicates the energy-flux-weighted centroid as given by Eq.~\eqref{eq: centroid_coord}, providing the best estimate of the main jet propagation direction relative to the injection axis. 
    The white regions at the top and bottom of each projection correspond to the regions excluded to avoid the polar axis singularity in the 3D simulations, as described in the second paragraph of Sect.~\ref{sec: initialdata}.}
    \label{fig: polar_plots}
\end{figure}

Fig.~\ref{fig: polar_plots} shows, for our three models, the polar ($\Theta,\Phi$) distributions of the energy per unit solid angle along with the position of the energy density flux centroids, 0.6\,s after the jet launch. While models A and C show a very good alignment with the original jet axis along $\vers{y}$, model B presents a remarkable offset of about 13 degrees. 

In the new spherical frame with polar direction $\vec{n}$, we can define the new polar angle $\Theta''$ as
\begin{equation}
   {\Theta}'' = \arccos(\vec{n}\cdot \vers{r}) \, .
\end{equation}
The new azimuthal angle $\Phi''$ is found by computing any orthonormal pair $(\vec{e}_1,\vec{e}_2)$ perpendicular to $\vec{n}$ with the Gram-Schmidt algorithm.
Vectors are projected on the new tilted basis, defined as:
\begin{equation}
    \hat{\bm\Phi}''=\dfrac{\vec{n}\times\vers{r}}{|\vec{n}\times\vers{r}|},  \quad \hat{\bm\Theta}''=\hat{\bm\Phi}''\times\vers{r} \, , 
\end{equation}
such that the original vector field $\vec{A}$ has components
\begin{equation}
    A''_{\Theta}=\vec{A}\!\cdot\!\hat{\bm\Theta}'', \quad
A_{\Phi}''=\vec{A}\!\cdot\!\hat{\bm\Phi}'' \, .
\end{equation}

We can then perform an axisymmetrization of the system by taking an energy density-weighted $\Phi''$-average of the different quantities around the new axis $\vec{n}$. Weighting on the local energy density, thus enhancing the contribution of the energetically dominant jet material, allows us to better preserve the total energy of the system, which remains the same within $\sim\!1\%$ (see Sect.~\ref{sec: axisym}).

Since models A and C are already well aligned with the original jet injection axis, we can perform the axisymmetrization around it, by averaging with respect to $\Phi$ (instead of $\Phi''$), without any significant difference. 
Only for model B, we need to perform the axisymmetrization around the new axis corresponding to the centroid defined above.

\subsection{Comparison between 3D and 2D evolution}
\label{appendix: 3D_2D_comparison}

\begin{figure}
    \centering
    \includegraphics[width=0.78\linewidth]{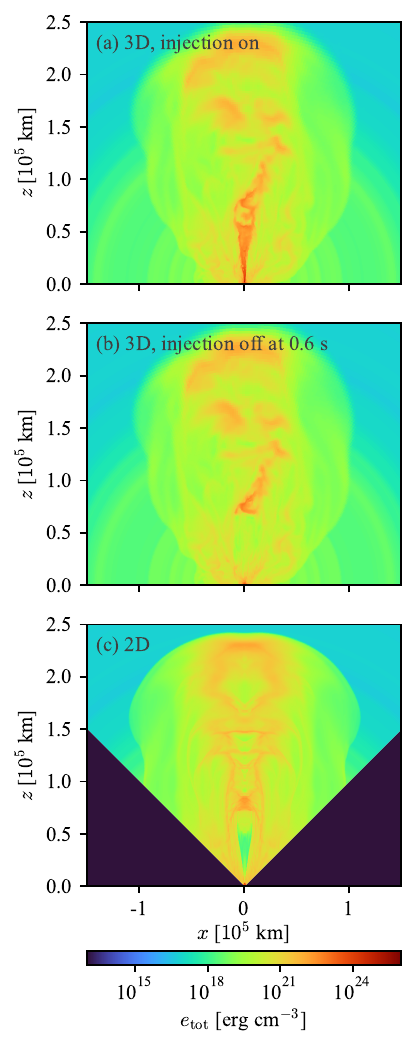}
    \caption{Comparison of the total energy density maps of three variants of model C, at 0.85 s after jet launch: the 3D run with continued injection (top), the 3D run with injection halted at 0.6~s (middle), and the 2D axisymmetrized run continuing the evolution after 0.6\,s (i.e.~the actual model C; bottom). The jet head position is the same in all three cases where the more blurred appearance of the shock front in the 3D run is due to the coarser (logarithmic scaling) grid resolution compared to the finer 2D grid.}
    \label{fig: injection_3D_2D}
\end{figure}

Here we assess the two main approximations of our pipeline: (i) switching off the central engine at $t-t_\jet=0.6$\,s, and (ii) continuing the evolution in axisymmetry after this time (Sect.~\ref{sec: axisym}).

Since in our setup the jet luminosity decays exponentially with timescale $\tau_L=0.3$\,s (see Sect.~\ref{sec: jetprescription}), $t-t_\jet=0.6$\,s corresponds to $2\,\tau_L$, implying that the instantaneous injected power has dropped by $e^{2}\!\approx\!7.4$ with respect to its initial value, while the energy still to be injected after this time is $e^{-2}\!\approx\!13.5\%$ of the total. 
This residual is non-negligible, but by $0.6$\,s the leading jet-cocoon front has already decoupled from the injection region, having broken out of the dense inner outflows and having started its propagation into the more tenuous dynamical ejecta; the residual injection feeds only the slow material near the base and does not reach the forward shock that sets the breakout.

To demonstrate this explicitly, we performed two additional 3D runs related to model~C, in which the evolution is continued beyond $0.6$\,s after jet launch with and without switching off the jet injection at that time.
Figure~\ref{fig: injection_3D_2D} compares, at the same absolute time $t-t_\jet=0.85$\,s (i.e.~$250$\,ms after the nominal engine switch off time), the total energy density $e_\tot$ in the meridional plane for the three variants of model~C: the 3D run with injection continued past $0.6$\,s (top), the 3D run with injection switched off at $0.6$\,s (middle), and the 2D axisymmetric continuation of the evolution after 0.6\,s (bottom).

The two 3D runs are indistinguishable at the leading front: the forward-shock radius and the jet-cocoon head coincide, and the only significant difference is an excess of energy confined to the innermost region ($r\lesssim 7\times10^4$\,km) of the injection-on run. The extra $\sim\!13.5\%$ of injected energy therefore remains near the base and never catches up with the front, confirming that the breakout radius, time, and the resulting GW-EM delay are insensitive to the fact that the injection is switched off at 0.6\,s.

The 2D axisymmetric continuation reproduces the jet-cocoon morphology and energy stratification of the 3D injection-off run, as well as the front position within grid spacing accuracy, showing that the energy-weighted azimuthal averaging of Sect.~\ref{sec: axisym} preserves the dynamics that matters for the light-curve calculation, while enabling the higher resolution and longer evolution required to reach breakout.

\section{Reconstruction and evolution of the dynamical ejecta}
\label{appendix: dynamical ejecta}
\begin{figure}
    \centering
    \includegraphics[width=0.98\linewidth]{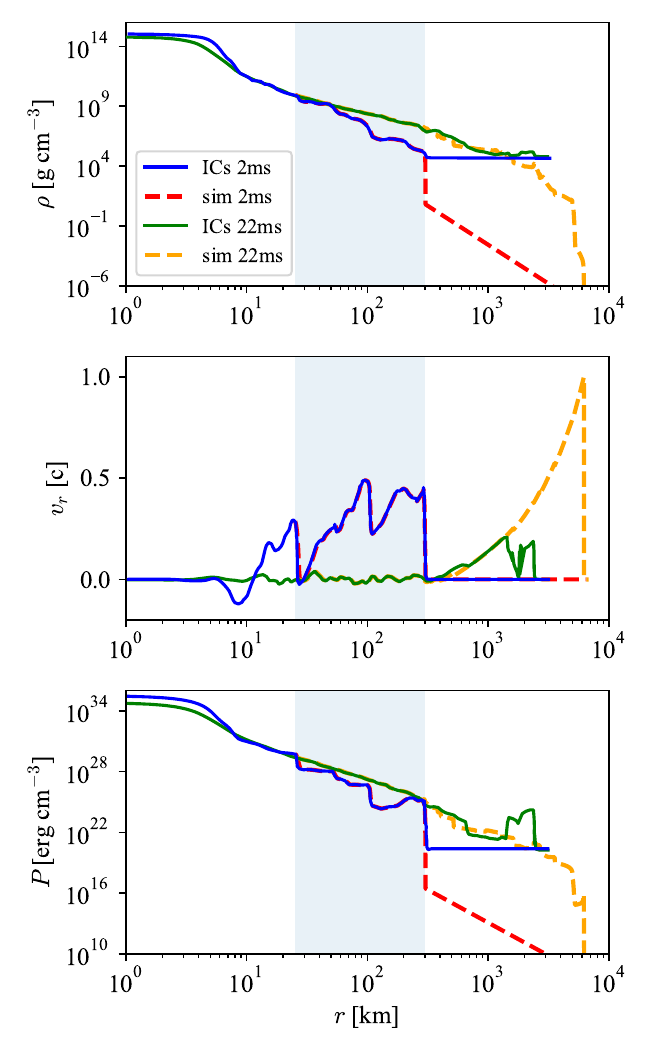}
    \caption{Radial profiles of rest-mass density (upper panel), radial velocity (middle panel), and pressure (lower panel). The blue and green lines are the original dynamical ejecta profiles at 2\,ms and 22\,ms after merger, respectively, while the red and yellow lines are the corresponding profiles reconstructed and simulated in \textsc{pluto}. 
    The shaded region, from the inner radius of the 1D evolution in \textsc{pluto} up to 300\,km, marks the radial interval where we substitute every 2\,ms the evolved profiles with the original ones from the BNS merger simulation (see text for details).}
    \label{fig: dynamical ejecta}
\end{figure}

\begin{figure}
    \centering
    \includegraphics[width=0.98\linewidth]{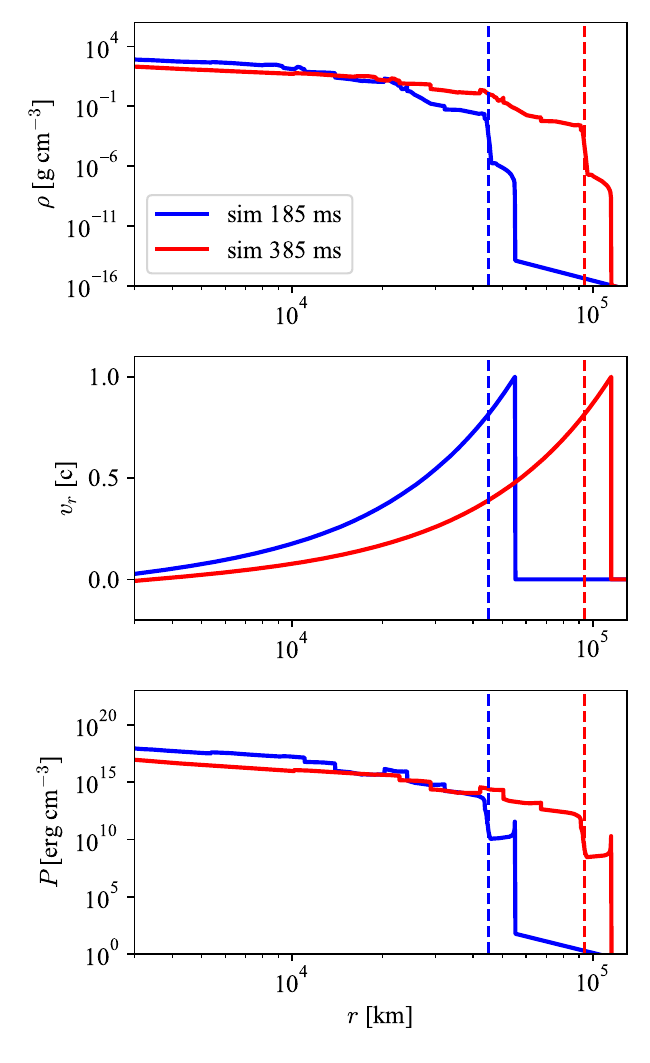}
    \caption{Same as Figure~\ref{fig: dynamical ejecta} but showing the reconstructed dynamical ejecta at 185 and 385\,ms after merger.
    The vertical dashed lines (at about $45000$ and $93000$\,km, respectively) mark the radii enclosing the bulk of these ejecta, where density drops suddenly by orders of magnitude. The mass outside these radii is negligible ($\simeq 10^{-9}\,\msun$).
    }
    \label{fig: ejecta 185ms and 385ms}
\end{figure}

To properly include the dynamical ejecta in our jet propagation environment, we proceed as follows:  
\begin{enumerate}
    \item We set up a 1D simulation in \textsc{pluto} starting from the radial profiles along the positive $z$-axis of rest-mass density, pressure, and radial velocity imported from the original BNS simulation at $t\!=\!2$\,ms after merger. The front of the dynamical ejecta, at such time, extends up to about $300~\km$ distance.

    \item For $r\!>\!300$\,km, we removed the uniform density floor ($\rho_\floor \simeq 6.3\times 10^4~\g/\cm^3$) and corresponding uniform pressure ($P_\floor \simeq 2.79\times 10^{20}~\erg/\cm^3$)
    of the merger simulation, and imposed `atmospheric' density and pressure profiles as 
    \begin{equation}
        \rho_\atmo(r) = \rho_\floor \left(\frac{r}{r_\rho}\right)^{-6.5}~\g~\cm^{-3},
    \end{equation}
    \begin{equation}
        P_\atmo(r) = P_\floor \left(\frac{r}{r_\rho}\right)^{-6.5}~\erg~\cm^{-3},
    \end{equation}
    where $r_\rho\!\simeq\!73.8$\,km (equal to 50 in geometrized units).
    This gives the red profiles in Figure~\ref{fig: dynamical ejecta} (substituting the original blue ones).

    \item We evolved the system in 1D for $\sim\!2$\,ms, using an inner radius of 25\,km (i.e.~the portion at $r\!<\!25$\,km is not included in the evolution) and applying a Paczy\'{n}ski--Wiita potential \citep{paczynsky_thick_1980} to account for the gravitational pull of the central $2.596~\msun$ black hole.

    \item After $2$\,ms of evolution, we substituted density, pressure, and radial velocity for $r\!<\!300~\km$ using the corresponding 1D profiles from the original BNS simulation at the same time. Meanwhile, the material at $r \geq 300~\km$ has expanded into the low-density `atmosphere', naturally developing a high velocity tail.

    \item We repeated step (4) every $2$\,ms up to $22$\,ms after merger, the time over which dynamical ejecta are produced. Note that the ejecta velocity at $r\!<\!300$\,km never exceeds 0.5\,$c$ and thus in 2\,ms there is no causal connection between the central object and the material at 300\,km distance. As a consequence, after any substitution the profiles remain perfectly smooth and continuous at 300\,km.  
    Figure~\ref{fig: dynamical ejecta} shows a comparison between the reconstructed dynamical ejecta at 22\,ms (yellow) after merger and the equivalent from the original simulation (green). The latter is clearly missing the fast outer part, lost into the high and uniform density floor. The reconstructed profiles are instead similar to those consistently obtained in \citet{kalinani_jet-environment_2025}, where BNS merger simulations are performed with very low density floor (in that case, decreasing as $r^{-6}$).
    
    \item We evolved the obtained 1D profiles of density, pressure and radial velocity from 22\,ms up to $385$\,ms after merger.
    The same profiles, taken either at 185 or at 385\,ms after merger, depending on the model under consideration, are then mapped into isotropic distributions in 3D spherical coordinates, to be finally employed to define the dynamical ejecta in our jet simulations (see Fig.~\ref{fig: ejecta 185ms and 385ms}). 
    We note that the total mass of this full, spherically reconstructed 1D dynamical-ejecta profile, integrated from the inner to the outer radii, is $0.03~\msun$.
    This refers to the isotropic reconstruction of the entire profile: once blended into the 3D environment, only the outer portion beyond the front of the post-merger outflow is tagged as `dynamical ejecta' (at $r\!\gtrsim\!1.8\times10^4$\,km for the $385$\,ms model, carrying $\simeq\!5\times10^{-4}~\msun$), while the inner part overlaps with, and is subsumed into, the post-merger outflow (`inner') component.
\end{enumerate}

\section{Detailed jet setup and energy calculation}
\label{appendix: jet}

\subsection{Jet luminosity}
\label{appendix: jet-lum}

We start from the stress-energy tensor of a relativistic, unmagnetized fluid
\begin{equation}
\label{eq: HD_tensor}
    T^{\mu\nu}_{\mathrm{HD}} = \rho h c^2 u^\mu u^\nu + P g^{\mu\nu} \, ,
\end{equation}
where $u^\mu = \Gamma(1,\vec{\beta})$ is the dimensionless 4-velocity with Lorentz factor $\Gamma$,
$g^{\mu\nu}$ is the metric tensor (which coincides with the Minkowski space-time $\eta^{\mu\nu}$ in our case), $h$ is the comoving dimensionless specific enthalpy defined as
\begin{equation}
    \label{eq: enthalpy}
    h = 1 + \dfrac{\gamma}{\gamma-1}\dfrac{P}{\rho c^2} \quad ,
\end{equation}
where $\rho$, $P$, and $\gamma$ are the rest-mass density, pressure, and adiabatic index, respectively. 
Since we use the Taub-Matthews EoS \citep{mignone_equation_2007}, pressure and density are linked by 
\begin{equation}
\label{eq: rhohc}
    \rho h c^2  = \dfrac{5}{2} P + \sqrt{\rho^2 c^4 + \dfrac{9}{4} P^2} \quad ,
\end{equation}
which gives
\begin{equation}
\label{eq: pressure_taub}
    P = \dfrac{1}{8} \left(5 h - \sqrt{9 h^2 + 16} \right) \rho c^2
    \approx \dfrac{h-1}{4} \rho c^2\quad ,
\end{equation}
where the last approximation is valid for hot relativistic jets. 

Since our jets are also magnetized, we have to include the electromagnetic contribution to the stress-energy tensor
\begin{equation}
\label{eq: EM_tensor}
    \begin{aligned}
        T^{\mu\nu}_{\mathrm{EM}} &= F^{\mu\alpha}F^\nu_{\;\ \alpha} - \dfrac{1}{4}\eta^{\mu\nu} F_{\alpha\lambda}F^{\alpha\lambda} \\ 
        &= b^2 u^\mu u^\nu + \dfrac{b^2}{2} g^{\mu\nu} - b^\mu b^\nu \quad , 
    \end{aligned}
\end{equation}
where $b^\mu = (b^0, \vec{b}\equiv b^i) $ represents the comoving magnetic field,\footnote{We employ a convention where factors $1/\sqrt{4\pi}$ are included in the magnetic field definition.} which is related to the lab frame magnetic field $\vec{B}$ via
\begin{equation}
    b^0 = \Gamma (\vec{\beta} \cdot \vec{B}) \quad , \quad  b^i = \dfrac{B^i}{\Gamma} + \Gamma (\vec{\beta} \cdot \vec{B})\beta^i \quad ,
\end{equation}
and its squared norm is
\begin{equation}
\label{eq: b_module}
    b^2 = \dfrac{B^2}{\Gamma^2} + (\vec{\beta} \cdot \vec{B})^2 \equiv B^2 - (\vec{\beta} \times \vec{B})^2 \quad ,
\end{equation}
with $B^2 = (B^r)^2 + (B^\theta)^2 + (B^\phi)^2$. 
From \eqref{eq: b_module} we can define the comoving magnetization $\sigma$ as
\begin{equation}
    \label{eq: magnetization}
    \sigma \equiv \dfrac{b^2}{\rho h c^2} \quad . 
\end{equation}
The total stress-energy tensor for a magnetized fluid is
\begin{equation}
\label{eq: total_stress_energy_tensor}
    T^{\mu\nu} = \rho h^* c^2 u^\mu u^\nu + \left( P + \dfrac{b^2}{2}\right) g^{\mu\nu} - b^\mu b^\nu \quad ,
\end{equation}
where we defined $h^* \equiv h ( 1 + \sigma) = h + b^2/\rho c^2$.
Magnetization and pressure can be expressed in terms of $h^*$ as
\begin{equation}
\label{eq: sigma_magnetic}
    \sigma = \dfrac{b^2}{\rho h^* c^2 - b^2 } \quad , 
\end{equation}
\begin{equation}
\label{eq: pressure_magnetic}
    P = \dfrac{h^*-1}{4} \rho c^2 - \dfrac{b^2}{4} \quad ,
\end{equation}
where we used the ultra-relativistic approximation for the latter formula.

The jet is injected with an axisymmetric structure along the radial direction through a polar cap at the radius $r_0$ and with half-opening angle $\theta_\jet$ and the 
corresponding one-sided jet luminosity is equal to 
\begin{equation}
    \label{eq: L_j1}
    \begin{aligned}
    L_\jet =2\pi r_0^2 \int_0^{\theta_\jet}& \left[ \rho_\jet h^*_\jet  c^2 \Gamma^2_{\jet}  \beta^r_{~\jet} \right. \\ & \left.- B^r_{~\jet} (\vec{\beta}_{\jet}\cdot \vec{B}_\jet) - \Gamma^2_{\jet} (\vec{\beta}_{\jet}\cdot \vec{B}_\jet)^2 \beta^r_{~\jet} \right] c \sin\theta\de\theta \ ,
    \end{aligned}
\end{equation}
where the subscript `$\jet$' indicates quantities associated with the jet injection.
The (uniform) jet's angular velocity $\Omega_{0,\jet}$ is set by averaging the angular velocity of the surrounding environment, measured at the jet edges.
This corresponds to a $\beta^\phi_{~\jet} \ll 1$ and thus $\Gamma_{\jet}$ is essentially given by the Lorentz factor along the radial direction.

\subsection{Transverse equilibrium}
\label{appendix: transverse_equilibrium}

Transverse equilibrium within the jet, which we impose at the injection radius $r_0$, corresponds to the balance between magnetic, thermal, and centrifugal forces acting along the direction of the polar angle $\theta$, i.e.~orthogonally to the radial and the azimuthal directions (in a spherical coordinate system aligned with the jet injection axis). 
For a steady flow, this condition is trivially expressed by $\nabla_j T^{\theta j} = 0$.
This translates to the following expression (assuming $\beta^\theta = 0$ and $B^\theta = 0$):
\begin{equation}
    \dfrac{1}{\sqrt{-g}} \di_\theta\left( \sqrt{-g}~ T^{\theta\theta}\right) + \Gamma^\theta_{\ \ \phi\phi}~T^{\phi\phi}= 0 \quad ,
\end{equation}
where $\sqrt{-g} = r^2 \sin\theta$ is the determinant of the metric tensor, while $\Gamma^\theta_{\ \ \phi\phi} = - \sin\theta\cos\theta$.
Using the total stress-energy tensor given in Eq.~(\ref{eq: total_stress_energy_tensor}), we get the following expression for the transverse equilibrium:
\begin{equation}
    \dfrac{\di}{\di \theta} \left( P + \dfrac{b^2}{2}\right) = \cot\theta\left[(\rho hc^2 + b^2)(\Gamma\beta^\phi)^2 - (b^\phi)^2 \right] \quad . 
\end{equation}
Here, $u^\phi = \Gamma\beta^\phi$ and $b^\phi$ represent the physically homogeneous azimuthal velocity and magnetic field components, respectively (i.e. the velocity component $u^i = \de X^i/c\de \tau$, with $X^i=(r,\theta,\phi)$, multiplied by its corresponding tensor weight $g_{ii}$).
From the relation between thermal pressure and density given by Eq.~\eqref{eq: pressure_magnetic} we find:
\begin{equation}
    \begin{aligned}
        \dfrac{\di \rho(\theta)}{\di \theta} &- \Gamma^2(\beta^\phi)^2 \dfrac{4h^*}{h^* -1} \cot(\theta) \rho(\theta) \\ &= -\dfrac{1}{(h^*-1) c^2}\left[ 4\cot\theta (b^\phi)^2 + \dfrac{\di b^2}{\di\theta}\right] \quad . 
    \end{aligned}
\end{equation}
The above first-order linear differential equation for $\rho(\theta)$ has the form $\rho'(\theta) - f(\theta)\rho(\theta) = -g(\theta)$, which admits the following exact solution: 
\begin{equation}
    \rho(\theta) = e^{\int_0^\theta f(\xi)\de\xi} \left[ \rho_0 - \int_0^\theta  g(\zeta) e^{-\int_{0}^\zeta f(\xi)\de\xi} \de\zeta\right] \quad ,
\end{equation}
where $\rho_0 \equiv \rho(\theta=0)$.

Applying the above formula at the jet injection radius $r_0$, the exact solution for the integral of $f(\xi)$ is
\begin{equation}
    \begin{aligned}
        \int_0^\theta f(\xi)\de\xi &= \dfrac{4h^*}{h^* - 1} \int_0^{\theta} \dfrac{(\beta^\phi_{~\jet})^2  \sin\xi\cos\xi}{1-(\beta^r_{~\jet})^2 - (\beta^\phi_{~\jet})^2 \sin^2\xi} \de\xi \\ 
        &= \dfrac{4h^*}{h^*-1} \ln\left( \dfrac{\Gamma_{\jet}}{\Gamma^r_{~\jet}}\right) \quad ,
    \end{aligned}
\end{equation}
where $\beta^\phi_{~\jet} = \Omega_{0,\jet} r_0/c$.
Since $\beta^\phi_{~\jet}\!\ll\!1$ and assuming that the jet radial velocity is ultra-relativistic, the total Lorentz factor nearly coincides with the radial one ($\Gamma_{\jet}\!\simeq\!\Gamma^r_{~\jet}$), implying $\ln ( \Gamma_{\jet}/\Gamma^r_{~\jet})\simeq 0$. Thus, we can further simplify our expression for $\rho(\theta)$ as
\begin{equation}
\label{eq: rho_theta}
    \rho(\theta) \simeq \rho_0 - \int_0^\theta g(\zeta) \de \zeta \quad .
\end{equation}
$\Gamma_{\jet}\!\simeq\!\Gamma^r_{~\jet}$ also implies
\begin{equation}
    b^\phi \simeq \dfrac{B^\phi}{\Gamma^r_{~\jet}} \quad , \quad 
    b^2 \simeq  \dfrac{(B^\phi)^2}{(\Gamma^r_{~\jet})^2} + (B^r)^2 \quad ,
\end{equation}
and, writing the lab-frame magnetic field components as 
$B^\phi = B_0 \lambda(\theta)$ and $B^r = B_0 \chi(\theta)$, the function $g(\zeta)$ in Eq.~\eqref{eq: rho_theta} becomes
\begin{equation}
    \begin{aligned}
        g(\zeta) = &\dfrac{B_0^2}{(h^*-1) c^2 (\Gamma^r_{~\jet})^2} \times \\
        &\left[ 4 \lambda^2(\zeta) \cot\zeta + \dfrac{\di[ \lambda^2(\zeta) + (\Gamma^r_{~\jet})^2 \chi^2(\zeta)]}{\di \zeta}\right] \, .
    \end{aligned}
\end{equation}
Using the above expression and plugging Eq.~\eqref{eq: rho_theta} into Eq.~\eqref{eq: L_j1} leads to the general solution for $\rho_0$, which reads
\begin{equation}
\label{eq: rho0_lambda_chi}
    \rho_0 = \dfrac{L_\jet}{\mathcal{A}_\jet (\Gamma^r_{~\jet})^2 h_\jet^* \beta^r_{~\jet} c^3} + \dfrac{B_0^2 \langle \Lambda(\theta_\jet)\rangle}{(\Gamma^r_{~\jet})^{2} c^2 (h_\jet^* -1)}  + \dfrac{B_0^2}{h_\jet^* c^2} \langle \chi^2(\theta_\jet) \rangle \, ,
\end{equation}
where $\mathcal{A}_\jet = 2 \pi r_0^2 (1-\cos\theta_\jet)$, and we defined 
\begin{equation}
\label{eq: lambda_average}
    \langle \Lambda(\theta_\jet)\rangle = \dfrac{1}{1-\cos\theta_\jet} \int_0^{\theta_\jet} \Lambda(\theta)\sin\theta\de\theta \, ,
\end{equation}
\begin{equation}
\label{eq: chi average}
    \langle \chi^2(\theta_\jet)\rangle = \dfrac{1}{1-\cos\theta_\jet} \int_0^{\theta_\jet} \chi^2(\theta)\sin\theta\de\theta \, ,
\end{equation}
with 
\begin{equation}
\label{eq: lambda_equation}
    \begin{aligned}
        \Lambda(\theta) = \int_0^\theta &4\lambda^2(\zeta)\cot\zeta~\de\zeta + \lambda^2(\theta)-\\
        & -\lambda^2(0) +(\Gamma^r_{~\jet})^2 [\chi^2(\theta)-\chi^2(0)] \, .
    \end{aligned}
\end{equation}

For the magnetic field functions, we adopt (see \citealt{marti_structure_2015,geng_propagation_2019})
\begin{equation}
    \lambda(\theta) = \dfrac{2 (\theta/\theta_\mathrm{m})}{1 + (\theta/\theta_\mathrm{m})^2}  \,\, ,
\end{equation}
\begin{equation}
    \chi(\theta)= 0.5 \,\, ,
\end{equation}
with $\theta_\mathrm{m} = 0.4\,\theta_\jet$.

Since we choose a small angle $\theta_\jet = 10^\circ$, it is possible to approximate the solution to an analytical form with very high precision.
To integrate Eqs.~\eqref{eq: lambda_equation}, \eqref{eq: lambda_average}, and \eqref{eq: chi average}, we substitute the trigonometric functions with their Taylor series and obtain 
\begin{equation}
 \langle \Lambda(\theta_\jet)\rangle = \int_0^{\theta_\jet} \dfrac{\Lambda(\tilde{\theta})\sin\tilde{\theta}}{1-\cos\theta_\jet}\de\tilde{\theta} \simeq \dfrac{3.1 \theta_\jet^2 - 0.419 \theta_\jet^4}{1-\cos\theta_\jet} \approx 6.2 \quad , 
\end{equation}
and $\langle\chi^2(\theta_\jet)\rangle = \chi^2 = 1/4$.

The solution for $\rho(\theta)$ in transverse equilibrium reads:
\begin{equation}
\label{eq: rho_theta_final}
    \rho(\theta) = \rho_0 - \dfrac{B_0^2}{(h^*-1) c^2(\Gamma^r_{~\jet})^2} \Lambda(\theta) \quad ,
\end{equation}
where
\begin{equation}
\label{eq: rho0}
    \rho_0 = \dfrac{L_\jet}{\mathcal{A}_\jet h^*_\jet (\Gamma^r_{~\jet})^2 \beta^r_{~\jet} c^3} + \dfrac{6.2B_0^2 }{(\Gamma^r_{~\jet})^2 c^2 (h^*_\jet-1)} + \dfrac{B_0^2}{4h_\jet^* c^2}  \quad , 
\end{equation}
and 
\begin{equation}
\label{eq: lambda_theta_full}
     \Lambda(\theta) \simeq  \left[\dfrac{8(\theta_\mathrm{m}^2+3) x^2}{3(x^2+1)} -\dfrac{8}{3} \theta_\mathrm{m}^2\log(1+x^2)+ \lambda^2(\theta)\right] \, ,
\end{equation}
with $x \equiv \theta/\theta_\mathrm{m}$.
\begin{figure}[t!]
    \centering
    \includegraphics[width=0.97\linewidth]{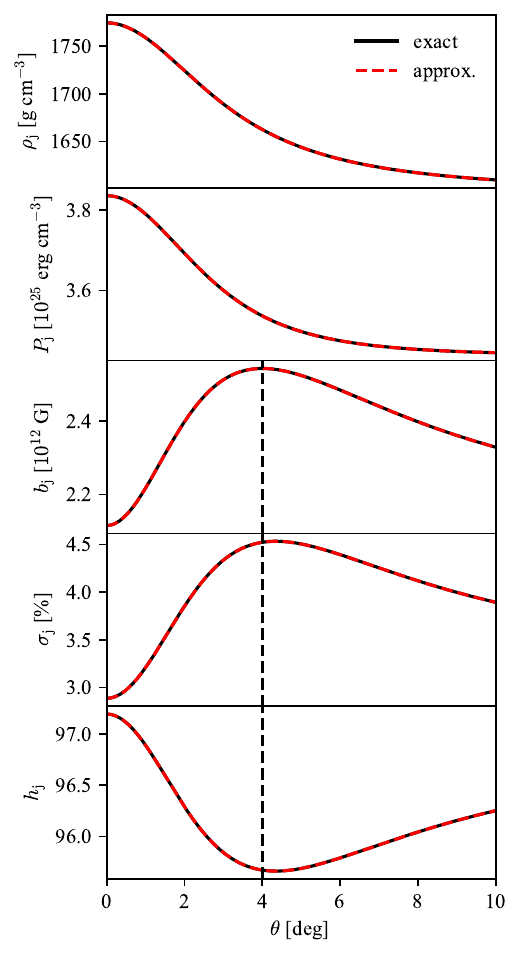}
    \caption{
        Polar angle profiles of jet injection quantities (density, pressure, comoving magnetic field, magnetization, enthalpy) obtained for our model C by imposing transverse equilibrium. The black lines show the direct integration, while the red dashed lines represent our analytical approximation (perfectly matching the exact result). The maximum variation of density and pressure across the angular profile is around $6\%$ and $12\%$, respectively, resulting in a nearly top-hat configuration. Vertical dashed line marks $\theta=\theta_\mathrm{m}$ (see text).
        }
    \label{fig: transverse_equilibrium}
\end{figure}

In Figure~\ref{fig: transverse_equilibrium}, we show the angular profiles of various jet injection quantities, referring in particular to our model C. From the comparison between the solution obtained with and without the small angle approximation $\theta_\jet \ll 1$, we see that differences are negligible.

\section{Shock finder algorithm}
\label{appendix: shock_finder}
We identified and flagged cells as shock-zones in our RMHD simulations by imposing the following conditions that a cell should meet at once \citep{gupta_numerical_2021}:

\begin{enumerate}
    \item converging flow with $\nabla\cdot \vec{v} < 0$ ;
    \item $\nabla T \cdot \nabla \rho > 0$ ;
    \item pressure jump between adjacent cells above a given threshold: $|\nabla P |/ \vec{P}_{\min,t} \cdot \Delta \vec{x} \geq \epsilon_\thresh $ .
\end{enumerate}
The temperature is defined in the comoving frame as

\begin{equation}
    T = \dfrac{\mw \mprot}{\kboltz} \dfrac{P}{\rho} \quad , 
\end{equation}
where $\mprot$ is the proton mass, $\mw$ is the mean molecular weight, and $\kboltz$ is the Boltzmann constant.
For this last condition, we considered a rather weak pressure jump threshold of $\epsilon_\thresh  = 10^{-3} - 10^{-2}$. 
The quantity $|\nabla P|/\vec{P}_{\min}$ is a vector with components along each direction given by the ratio between the gradient of pressure in absolute value and the minimum of pressure among adjacent cells, while $\Delta \vec{x}$ is the grid spacing vector (see \citealt{mignone_pluto_2011}). 

\section{Light-curve calculation}
\label{appendix: luminosity_curves}

\subsection{Shock breakout emission}

Our treatment of the shock-breakout emission follows the relativistic shock-breakout and cocoon-emission framework of \citet{nakar_early_2010, nakar_relativistic_2012}, and \citet{nakar_electromagnetic_2020}, adapted here to the angle-dependent, laterally expanding jet-cocoon shock. 
Similarly to \citet{gottlieb_cocoon_2018}, the light curve calculation is performed in post-processing from the internal energy stored in the breakout layer of the RMHD simulation, released as the photosphere recedes in time, and corrected for light-travel time and relativistic beaming.
For each angle and time, we tracked the shock position with the procedure described in Appendix~\ref{appendix: shock_finder}.
For a fixed angular direction $\theta_\obs$, we define the shock-breakout time $t_\bo$ as the time at which the shock reaches the radius $r_\bo$ satisfying the condition 
$\tau \simeq 1/\beta_\sh'$ (see Eq.~\eqref{bo-condition}).
After this time, photons trapped within the shock start to be released from the emission radius $r_\mathrm{e}(t)$, corresponding, at each given time, to the radius where $\tau \simeq 1/\beta_\sh'$.

Assuming a spherically symmetric system with the same properties we have along $\theta_\obs$, we computed the comoving-frame thermal energy released in a given simulation timestep $\Delta t$ between $t_1$ and $t_2$ as that contained, at $t_2$, between $r_\mathrm{e}(t_2)$ and $r_\mathrm{e}(t_1)+v_\mathrm{e}(t_1) \Delta t$, where $v_\mathrm{e}$ is the fluid velocity at the emission radius \citep{gottlieb_cocoon_2018, Gutierrez2025},
\begin{equation}
   \Delta E' (t_1|t_2) \simeq 4\pi \int_{r_\mathrm{e}(t_2)}^{r_\mathrm{e}(t_1)+v_\mathrm{e}(t_1) \Delta t} e'_\mathrm{th}(t_2)\,\Gamma(t_2)\, r^2 \de r \, , 
   \label{comEn_radial}
\end{equation}
where $e'_\mathrm{th}\!\simeq\! 3P$ is the comoving thermal energy density. 
Considering the comoving timestep $\Delta t'\!=\!\Delta t/\Gamma$ (where $\Gamma$ here is an average value for the emitting layer's Lorentz factor in the given timespan), we can write the comoving energy release per unit time as $\Delta E'/\Delta t'$, which represents an average luminosity in the time interval from $t_1$ and $t_2$. We take this quantity as our first estimate of the comoving, isotropic-equivalent bolometric luminosity of the shock breakout emission ($\mathcal{L}'_\mathrm{iso}$). 
In the following, we refine this estimate to account for non-radial emission.

\subsection{Including non-radial contributions}

Since photon diffusion is not purely radial, the observer will receive radiation from different portions of the emitting layer, with each portion moving along a direction that is in general misaligned with respect to the observing angle.
In order to collect radiation from those different portions, we need to account for different light travel times as well as non-radial optical paths.

Being the observer at very large distance compared to the size of the system, the rays connecting the emitting layer with the observer are essentially parallel to each other. 
Let $\hat{\vec{n}}_\obs = (\sin\theta_\obs,0,\cos\theta_\obs)$ be the unit vector pointing towards an observer at $\theta=\theta_\obs$ (and, for simplicity, $\phi_\obs=0$), and $\hat{\vec{n}} = (\sin\theta\cos\phi,\sin\theta\sin\phi,\cos\theta)$ the normal vector to the spherical surface at given $(\theta,\phi)$. 
The relative angle $\psi$ between the two unit vectors is given by the following expression:
\begin{equation}
    \mu \equiv \cos\psi\equiv \hat{\vec{n}}_\obs \cdot \hat{\vec{n}} 
    = \sin\theta_\obs\sin\theta \cos\phi + \cos\theta_\obs\cos\theta \,.
\end{equation}

We first restrict to the meridional plane containing the observer's location. 
\begin{figure}
    \centering
    \includegraphics[width=0.98\linewidth]{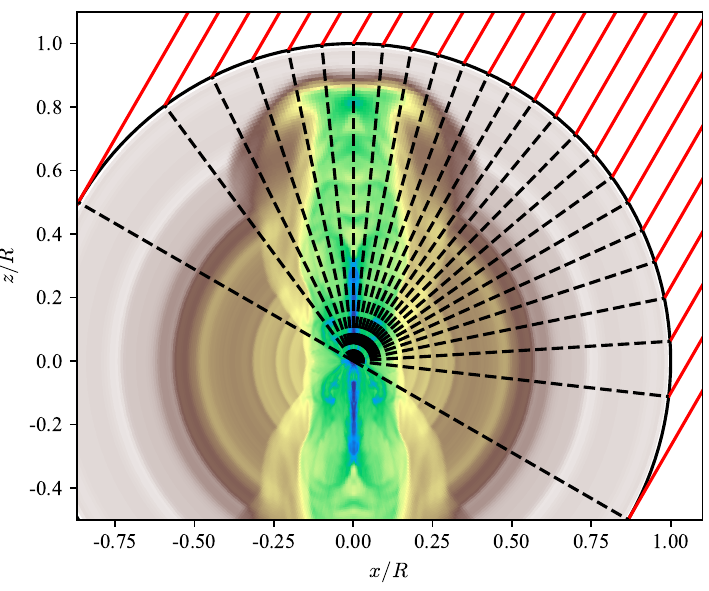}
    \caption{
        2D section of the total energy density of model~A at $t-t_\jet = 0.6~\mathrm{s}$ superposed to a pencil of parallel lines connecting to an observer at $\theta_\obs=30^\circ$ (red lines). The black dashed lines show the corresponding angle at which the rays hit the spherical surface of the dynamical ejecta front, determining their relative inclination. Note that the southern hemisphere is added (copying from the northern one) to cover the full radial extension of the dynamical ejecta front, with radius $R=1.9\times10^5~\km$ at that specific moment. 
        }
    \label{fig: non_radial_contribution}
\end{figure}
In this case, $\mu = \cos(\theta-\theta_\obs)$. 
At each evolution time, we considered a pencil of parallel and equally spaced rays oriented towards the observer and covering the full radial extension of the dynamical ejecta front, which is a spherical surface of radius $R$, as shown in Figure~\ref{fig: non_radial_contribution}. While we keep the number of lines fixed to 101 (central one aligned with $\theta_\obs$, plus $N=50$ for each side), the spacing $\lambda = R/N$ scales according to the dynamical ejecta expansion. 
The parametric equation describing the pencil of rays is
\begin{equation}
\label{eq: z_xn}
    z(x,n) = \frac{1}{\sin\theta_\obs} \left[n\lambda + x\cos\theta_\obs\right] 
    \quad \forall\,n \in \left[-N,N\right] \, ,
\end{equation}
where $n$ is an integer running from $-N$ to $N$.
It can be shown that the polar angle $\theta$ at which a ray shifted by $n\lambda$ with respect to the line of sight hits the spherical surface of radius $R$ is
\begin{equation}
\label{eq: theta_n}
    \theta_n = \theta_\obs - \arcsin\left(\frac{n}{N}\right) \, ,
\end{equation}
where positive $n$ values correspond to rays above the $n=0$ line ($\theta=\theta_\obs$), while negative $n$ correspond to rays below the line of sight.

At this point, we extend the above construction to cover also the $y$ direction: the collection of rays on the $xz-$plane (i.e. $y=0$) becomes a 2D pencil of $(2N+1)^2$ rays, where we maintain the same $N$ and $\lambda$ also along the $y$ direction. 
One specific ray is then identified by the integer $n$ (see above) and by the new integer $m\in \left[-N,N\right]$, which corresponds to the plane $y=m\lambda$.

\begin{figure}
    \centering
    \includegraphics[width=1.\linewidth]{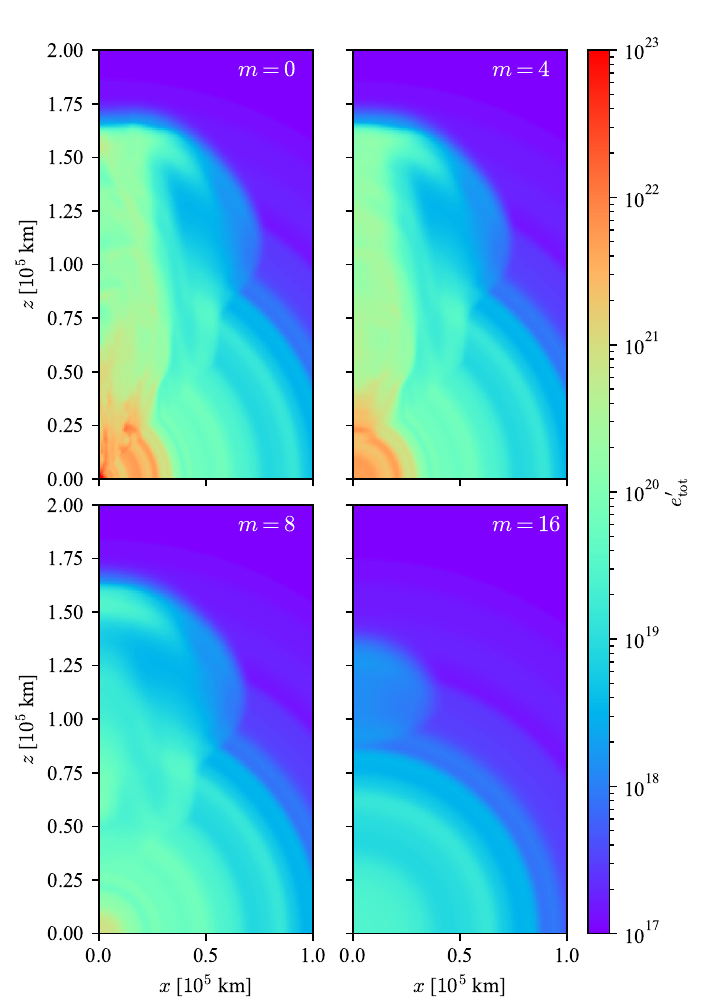}
    \caption{
        Total energy density of model~A at $t-t_\jet = 0.6~\mathrm{s}$, as seen on the planes $y=m\lambda$, with $\lambda=3.8\times 10^3~\km$ and $m=0,4,8,16$.
    }
    \label{fig: tomography}
\end{figure}

Since we evolve assuming axisymmetry, we only had data on the $y\!=\!0$ plane and needed to reconstruct the equivalent information on $y\!\neq\!0$ planes.
This was done by maintaining fixed the $z$ coordinate and by defining a new $\chi$ coordinate as $\chi(m) = \sqrt{x^2 - (m\lambda)^2}$.
In Figure~\ref{fig: tomography} we illustrate an example of the resulting 2D slices for $m=0,4,8,16$, referring to model~A at $t-t_\jet = 0.6~\mathrm{s}$. 

The emitting region, projected on a plane orthogonal to the line of sight, can now be divided into $(2N+1)^2$ equal squares centred on each ray of the 2D pencil, identified via the integers $(m,n)$. 
For a given timestep of the evolution $\Delta t=t_2-t_1$, the contribution to the shock-breakout emission coming from one of those $(2N+1)^2$ squares will be associated with the comoving-frame thermal energy contained between the emission radius along the specific ray at $t_2$, indicated with $r_\mathrm{e}^{m,n}(t_2)$, and the radius at which the surface $r_\mathrm{e}(t_1,\theta,\phi)+v_\mathrm{e}(t_1,\theta,\phi) \Delta t$ intersects this ray, which is approximately given by 
$r_\mathrm{e}^{m,n}(t_1)+v_\mathrm{e}^{m,n}(t_1) \Delta t$. 
Naming $\tilde{x}^{m,n}$ the position along the chosen ray, we call $\tilde{x}_\mathrm{in}^{m,n}$ and $\tilde{x}_\mathrm{out}^{m,n}$ the two positions corresponding to the above radii. 
The resulting comoving-frame thermal energy can be computed as\footnote{Here, we are neglecting the time variation of $\lambda$ in a timestep $\Delta t$, since in our calculations this corresponds to a relative change of order $10^{-5}$.}
\begin{equation}
   [\Delta E' (t_1|t_2)]^{m,n} \simeq \lambda(t_2)^2   \int_{\tilde{x}_\mathrm{in}^{m,n}}^{\tilde{x}_\mathrm{out}^{m,n}} e'_\mathrm{th}(t_2)\,\Gamma_\perp(t_2)\Gamma_\parallel(t_2)\, \de \tilde{x}^{m,n} \,, 
   \label{comEn_radial_final}
\end{equation}
where we used the volume transformation $\de V' \!=\! \de S' \de \tilde{x}' \!=\!  \Gamma_\perp \de S~\Gamma_\parallel \de \tilde{x} \!=\! \Gamma_\perp\Gamma_\parallel \de V$, with $\Gamma_\parallel \!=\! \sqrt{\Gamma^2\mu^2 + (1-\mu^2)}$ and $\Gamma_\perp \!=\! \sqrt{\mu^2 + \Gamma^2(1-\mu^2)}$\, . 

We note that, in this case, the emission radii are obtained by integrating the optical depth along $\tilde{x}$ \citep{abramowicz_appearance_1991}:
\begin{equation}
    \tau = \int \kappa \rho(\tilde{x})\,\Gamma(\tilde{x}) [1-\mu\beta(\tilde{x})] \de \tilde{x} \, .
\end{equation}
A criterion analogous to $\tau \!\simeq \! 1/\beta_\sh'$ is still adopted, but now considering the motion along the line of sight, i.e.~$\tau \simeq 1 / \mu'\beta_\sh'= (1-\mu\beta_\ej) / [ \beta_\sh' (\mu-\beta_\ej)]$\,.\footnote{For increasing misalignment with the observer's direction, this expression would eventually diverge as $\mu$ approaches $\beta_\ej$. Nonetheless, radiation is already suppressed by relativistic beaming at a smaller misalignment, when $\mu$ equals the velocity in units of $c$ of the emitting layer, which is larger than $\beta_\ej$\,.} 

\subsection{Doppler effect and final luminosity}

Since the emission layer is moving at relativistic speed (with a certain Lorentz factor $\Gamma$ and corresponding velocity $\beta c$), the resulting radiation is affected by a relativistic Doppler shift in the direction of the observer that causes the aberration of light, such that the luminosity per unit solid angle $\de \mathcal{L}_\mathrm{obs}/\de \Omega$ measured by a stationary observer relates to the comoving-frame luminosity per unit solid angle as $\de \mathcal{L}_\mathrm{obs}/\de \Omega\propto \mathcal{D}^4 \, \de \mathcal{L}'/\de\Omega'$ \citep{rybicki_radiative_1979}, with the Doppler factor $\mathcal{D}$ given by 
\begin{equation}
    \mathcal{D} = \dfrac{\de t'}{\de T_\obs}\equiv \dfrac{1}{\Gamma(1-\mu\beta)} \quad \, ,
\end{equation}
where $t'$ is the proper time in the comoving frame.

Considering Eq.~(\ref{comEn_radial_final}) and applying the Doppler shift as above, the observed average bolometric luminosity per unit solid angle due to the portion ($m,n$) of the emitting layer between the evolution times $t_1$ and $t_2$ can be written as 
\begin{equation}
    \bigg{[}\frac{\de \mathcal{L}_\mathrm{obs}(t_1|t_2)}{\de \Omega}\bigg{]}^{m,n} = \mathcal{D}^4\frac{[\Delta E' (t_1|t_2)]^{m,n}}{4\pi \Delta t'} \quad \, ,
\label{dLdOmega}
\end{equation}
where $\Delta t'=\Delta t/\Gamma$\, .

The next key element to consider is the arrival time of the emitted photons from the point of view of the observer. Radiation coming from a position ($r,\theta,\phi$) at a time $t$ after merger, will arrive at 
\begin{equation}
    T_\obs(t,r,\theta,\phi) = t - \mu\,\frac{r}{c} \, ,
\label{eq: observer_time}
\end{equation}
where $T_\obs$ is the time since the observed peak GW signal associated with the BNS merger.  
Based on this, a contribution to the signal similar to the one given in Eq.~(\ref{dLdOmega}) will have an average arrival time $T_k^{m,n}$, where the index $k$ indicates the specific evolution timestep under consideration.

Dividing $T_\obs$ in bins of width $\Delta T$, we can define a discrete set of times $T_j$ corresponding to the centre of each bin via the running index $j$. At this point, we introduce a window function $\mathcal{W}(T_j - T_k^{m,n})$ that is equal to 1 when $T_k^{m,n}$ falls within the bin centred in $T_j$ (i.e.~$|T_j - T_k^{m,n}|<\Delta T/2$) and equal to 0 otherwise. 

As final step, we reconstructed the full observed bolometric luminosity per unit solid angle as function of the observer time $T_j$ by summing up the contribution from each portion $(m,n)$ of the emitting layer and for each timestep:
\begin{equation}
      \frac{\de \mathcal{L}_\mathrm{obs}}{\de \Omega}(T_j) = \sum_k\sum_{m,n=-N}^{N}  \bigg{[}\frac{\de \mathcal{L}_\mathrm{obs}}{\de \Omega}\bigg{]}_k^{m,n} \mathcal{W}(T_j - T_k^{m,n}) \, .
\end{equation}
Multiplying the above expression by $4\pi$ gives the isotropic-equivalent bolometric luminosity $\mathcal{L}_\mathrm{iso}(T_j)$.

We note that, in the above calculation, $\Delta T$ should be much larger than the evolution timestep $\Delta t$ and, at the same time, small enough to capture the relevant features of the evolving luminosity. Our choice of $\Delta T$ satisfies both requirements.  

\end{appendix}

\end{document}